\documentclass[seceq]{ptptex}

\usepackage{graphicx}

\usepackage{wrapft}

\newcommand {\beq}{\begin{equation}}
\newcommand {\eeq}{\end{equation}}
\newcommand {\bea}{\begin{eqnarray}}
\newcommand {\eea}{\end{eqnarray}}
\newcommand {\nn}{\nonumber \\}
\newcommand {\Tr}{{\rm Tr\,}}

\newcommand {\e}{{\rm e}}

\newcommand {\m}{\mu}
\newcommand {\n}{\nu}
\newcommand {\pl}{\partial}

\newcommand {\vp}{\varphi}

\newcommand {\al}{\alpha}
\newcommand {\be}{\beta}
\newcommand {\ga}{\gamma}

\newcommand {\La}{\Lambda}

\newcommand {\sh}{\theta}   

\newcommand {\om}{\omega}

\newcommand {\ep}{\epsilon}

\newcommand {\del}  {\delta}
\newcommand {\Del}  {\Delta}
\newcommand {\mn}{{\mu\nu}}

\newcommand {\half}{ {\frac{1}{2}} }

\newcommand {\fourth} {\frac{1}{4} }

\newcommand {\sqpp} {\sqrt{p^2}}

\newcommand {\Acal}{{\cal A}}
\newcommand {\Ecal}{{\cal E}}

\newcommand {\Lcal}{{\cal L}}

\newcommand {\Dcal}{{\cal D}}

\newcommand {\Vcal}{{\cal V}}
\newcommand {\Wcal}{{\cal W}}

\newcommand {\Ftil}{{\tilde F}}

\newcommand {\ptil} {{\tilde p}}
\newcommand {\ktil} {{\tilde k}}

\newcommand {\Fhat}{{\hat F}}

\newcommand {\What}{{\hat W}}

\newcommand {\delh} {{\hat \delta}}

\newcommand {\gh}  {{\hat g}}

\newcommand {\rdot}{\dot{r}}
\newcommand {\rddot}{\ddot{r}}
\newcommand {\udot}{\dot{u}}

\newcommand {\bfZ} {{\bf Z}}

\newcommand {\intfx} {{\int d^4x}}

\newcommand {\intxy} {{\int d^4xdy}}
\newcommand {\intxz} {{\int d^4xdz}}

\newcommand {\intp} {{\int \frac{d^4p}{(2\pi)^4}}}

\newcommand {\intt} {{\int_{0}^{\infty}\frac{dt}{t}}}
\newcommand {\change} {\leftrightarrow}
\newcommand {\ra} {\rightarrow}

\newcommand {\pr}   {{\quad .}}
\newcommand {\com}  {{\quad ,}}
\newcommand {\q}    {\quad}

\newcommand {\nl}    {\newline}

\newcommand {\npl}  {{\frac{n\pi}{l}}}
\newcommand {\nplz}  {{\frac{n\pi}{l_0}}}

\newcommand {\zpl}  {{\frac{z\pi}{l}}}
\newcommand {\zplz}  {{\frac{z\pi}{l_0}}}

\newcommand {\snbar} {{\bar{\mbox{sn}}}}
\newcommand {\cnbar} {{\bar{\mbox{cn}}}}
\newcommand {\snbarp} {{\bar{\mbox{sn}}}_+}
\newcommand {\snbarm} {{\bar{\mbox{sn}}}_-}
\newcommand {\cnbarp} {{\bar{\mbox{cn}}}_+}
\newcommand {\cnbarm} {{\bar{\mbox{cn}}}_-}
\newcommand {\snbarpm} {{\bar{\mbox{sn}}}_\pm}
\newcommand {\cnbarpm} {{\bar{\mbox{cn}}}_\pm}

\pubinfo{Vol.~121, No.~4, April? 2009}

\title{
Casimir Energy of 5D Electromagnetism and 
New Regularization Based on Minimal-Area Principle
}

\author{
Shoichi \textsc{Ichinose}\footnote{E-mail: ichinose@u-shizuoka-ken.ac.jp}
}

\inst{
Laboratory of Physics, School of Food and Nutritional Sciences, 
University of Shizuoka, \\
Shizuoka 422-8526, Japan
}

\recdate{
November 8, 2008; revised ????? ?, 2009}

\abst{
We examine the Casimir energy of 5D electromagnetism 
from the recent standpoint.  
The bulk geometry is flat. The Z$_2$ symmetry  and periodic property, 
for the extra coordinate, are taken into account. 
After confirming the consistency with the past result, 
we do new things based on a {\it new regularization}. 
In the treatment of the divergences, we introduce IR and UV cutoffs
and {\it restrict} the (4D momentum, extra coordinate)-integral region. 
The regularized configuration is the {\it sphere lattice}, in the 4D continuum 
space, which changes along the extra coordinate. 
The change (renormalization flow) is specified by the {\it minimal area principle}; hence, 
this regularization configuration is stringlike.  
We do the analysis not in the Kaluza-Klein expanded form but in a
{\it closed} form. We do {\it not} use any perturbation. 
The formalism is based on the heat-kernel
approach using the {\it position/momentum propagator}. 
Interesting relations between the heat kernels and the P/M propagators 
are obtained, where we introduce the {\it generalized}
P/M propagators. 
A useful expression of the Casimir energy, in terms of the P/M propagator, 
is obtained. 
The restricted region approach is replaced by the weight-function approach
in the latter-half description. Its meaning, in relation to {\it space-time quantization}, is argued. 
{\it Finite} Casimir energy is numerically obtained. The compactification-size parameter 
(periodicity) suffers from the renormalization effect. Numerical evaluation is exploited. 
In particular, the minimal surface lines in the 5D flat space are obtained both numerically 
using the Runge-Kutta method and analytically using a general solution. 
}

\begin{document}
\maketitle

\section{Introduction\label{S.Intro}}
As a unified theory of the four forces in nature, higher-dimensional
models have a long history since the papers by Kaluza\cite{Kal21} and 
Klein\cite{Klein26}. The simplest
one unifies the forces of gravitons, photons and dilatons. The quantum effects
are evaluated by Appelquist and Chodos\cite{AC83}. They evaluated Casimir
energy\footnote{
For a recent review, see Ref.\citen{CasBMM01}. 
}
 and the result has been giving us a standard image of the contraction
of the extra space, that is, when compactification takes place, 
the extra space shrinks to a size 
slightly larger than the Planck length. 

Higher-dimensional models have, at present, the defect that they, by themselves, 
are {\it unrenormalizable}. In Ref.\citen{AC83}, UV divergence appears
as quintic divergence $\La^5$ of the {\it cosmological term}. It simply means that 
we have no appropriate procedure for defining physical quantities within the quantum field theory (QFT).
One can, at this point, have the standpoint that they are effective theories
that should be 
derived  from more fundamental models such as the string theory, M-theory, and D-brane. 
In the present study, we pursue the 
possibility that there is an appropriate procedure for defining 
physical quantities within the higher-dimensional QFT. 
We propose a procedure and show that it works well. 

There are some approaches to solving the above problem. 
One of them is the deconstruction model\cite{ACGprl01,HPWpr01}. 
We discretize the extra coordinate and choose an appropriate finite number of ``branes" keeping
gauge invariance. This is
a commonly used approach at present. Some interesting results are reported\cite{BLS03,RST05}. 
The other approach is based on the regularization using 
a position-dependent cutoff\cite{RS01}. The integral region is restricted appropriately. 
The restriction requirement comes from the analysis of propagator behaviour\cite{RS01,IM0703}. 
Spiritually, the {\it holograpy} idea is behind the restriction procedure. 
The present motivation comes from 
the question, ``Can we find the reason why the restriction process is necessary 
within the framework of the 5D QFT, not using the string theory and related supergravity theories?"

We introduce a {\it new} regularization inspired by the partial success of Randall-Schwartz's
result. We associate the regularization (in 4D world) cutoffs running along the extra axis $y$, 
with the {\it minimal area surfaces} in the bulk. In this way, the stringlike (surface) configuration (closed string)
is introduced in the present approach. This is quite in contrast to the usual 
string theory approach. The present stringlike configuration appears 
not from the propagation of strings but 
from the necessity of the restriction of the
integral region in the bulk space.

The original approach to the renormalization flow interpretation of 
bulk behaviour relies on the AdS/CFT and 5D supergravity 
\cite{FMMRnp9804,HSjh9806,DZatmp9810,GPPZjh9810,PSpl9903,FGPWatmp9904}. 
The present approach does {\it not} rely on them.  
We {\it directly} use the minimal area principle, the essence of the string theory \cite{Nambu70,Goto71,Pol81}, 
in the {\it regularization procedure}.  This is {\it new} in the development 
of the quantum field theory.  

This paper is organized as follows. In \S 2, we review the 5D quantum electromagnetism
in the recent standpoint. Casimir energy is obtained from the KK-expansion approach. 
In \S 3, the same quantity of \S 2 is dealt with in the heat-kernel method, and 
Casimir energy is expressed in a closed form in terms of the P/M 
propagator. The closed expression of Casimir energy is numerically evaluated 
and its equivalence with the result of \S 2 is confirmed in \S 4. Here we 
introduce UV and IR regularization parameters in the(4D momentum, extra coordinate) space. 
A new idea about UV and IR regularization is presented in \S 5. The minimal surface
principle is introduced. The {\it sphere lattice} and {\it renormalization} are explained. In \S 6, 
an improved regularization procedure is presented where a {\it weight function} is 
introduced. Here, again the {\it minimal surface principle} is taken. The definition of the 
weight function is given in \S 7. In \S 8, we present the conclusions.   We prepare 
two appendices to supplement the text. Appendix~A  deals with the analytic solution of the minimal surface curve in the 
5D flat space. Appendix~B provides an explanation of the numerical confirmation of the (approximate) 
equality of the minimal surface curve and the dominant path in Casimir energy calculation.

\section{Five-dimensional quantum electromagnetism\label{5dEM}}
We consider the flat 5D space-time $(X^M)=(x^\mu,y)$ with the periodicity
in the extra space $y$, 
\bea
ds^2=\eta_\mn dx^\m dx^\n+dy^2\com\q
-\infty < y < \infty\com\q
y\ra y+2l,\nn
(\eta_\mn)=\mbox{diag}(-1,1,1,1)\ ,
(X^M)=(x^\m,x^5=y)\equiv (x,y)\ ,\nn 
M,N=0,1,2,3,5;\ \m,\n=0,1,2,3.
\label{5dEM1}
\eea
The 5D electromagnetism is described by the 5D U(1) gauge field
$A_M$, 
\bea
S_{EM}=\intxy \sqrt{-G}\{-\fourth F_{MN}F^{MN}\}\equiv\intxy\Lcal_{EM}\com\q
G=\mbox{det}~G_{MN}\com\nn
F_{MN}=\pl_M A_N-\pl_N A_M\com\q (X^M)=(x^\m,y)\com\nn
ds^2=G_{MN}dX^MdX^N\com\q (G_{MN})=\mbox{diag}(-1,1,1,1,1)
\label{5dEM2}
\eea
It has $U$(1) gauge symmetry, 
\bea
A_M\ra A_M+\pl_M\La\com
\label{5dEM3}
\eea
where $\La(X)$ is the 5D gauge parameter. 

We respect $Z_2$ symmetry in the extra space, 
\bea
y\ra -y
\pr
\label{5dEM4}
\eea
The $Z_2$ parity assignment of $A_M(x^\m,y)$ is
fixed by the 5D gauge transformation (\ref{5dEM3}). 
There are two cases corresponding to the choice of 
the $Z_2$ parity of $\La(x^\m,y)$, 
\bea
\mbox{Case 1.\ Even-parity case}\q \La(x^\m,y)=+\La(x^\m,-y)\nn
A_\m\ :\ P=+\com\q A_5\ :\ P=-\com\nn
\mbox{Case 2.\ Odd-parity case}\q \La(x^\m,y)=-\La(x^\m,-y)\nn
A_\m\ :\ P=-\com\q A_5\ :\ P=+
\pr
\label{5dEM5}
\eea
In the present paper, we consider Case 1. (Case 2 can be similarly treated.)

We take the following gauge-fixing term to quantize
the present system. 
\footnote{
The gauge independence of the physical quantities is an important 
check point of the proposal in the present paper. We relegate it 
to a future work. The gauge independence of Casimir energy 
of the 5D KK theory (Appelquist and Chodos's result\cite{AC83}) was confirmed in Ref. \cite{SI85PLB}. 
}
\bea
\Lcal_g=-\half (\pl_MA^M)^2=-\half (\pl_\m A^\m+\pl_yA^5)^2\com\nn
\Lcal_{EM}+\Lcal_g=\half A_\m (\pl^2+\pl_y^2)A^\m
+\half A^5(\pl^2+\pl_y^2)A^5+\mbox{total derivatives}
\com
\label{5dEM6}
\eea
where $\pl^2\equiv\pl_\m\pl^\m$. Then the field equations are given by
\bea
(\pl^2+\pl_y^2)A^\m=0\com\q (\pl^2+\pl_y^2)A^5=0
\pr
\label{5dEM7}
\eea
We consider the system in the periodic condition(\ref{5dEM1}). 
Then we can write $A^M$ as
\bea
A^\m(x,y)=a_0^\m(x)+2\sum_{n=1}^\infty a^\m_n(x)\cos\npl y\com\q
\mbox{P=+}\nn
A^5(x,y)=2\sum_{n=1}^\infty b_n(x)\sin\npl y\com\q
\mbox{P=}-
\com
\label{5dEM8}
\eea
where \{$a_n^\m(x)$\} and \{$b_n(x)$\} are the KK-expansion 
coefficients. From Eq.(\ref{5dEM7}), they satisfy
\bea
\pl^2a_0^\m=0\ \ (\mbox{zero mode}),\ \{\pl^2-(\npl)^2 \}a_n^\m=0,\ 
\{\pl^2-(\npl)^2\}b_n=0,\ n\neq 0
\ .
\label{5dEM8b}
\eea
The on-shell condition, Eq.(\ref{5dEM7}) or Eq.(\ref{5dEM8b}), is for the analysis
of the S-matrix. In this study, we do not use the condition.
\footnote{  
We do not take into account the degree of freedom 5$-$2=3 
among five components $\{A^M\}$ due to the local gauge symmetry. This is because 
we will compare the present results of the flat geometry with those 
of the warped geometry. In the latter treatment, 
we start with the {\it massive} vector theory that has no 
local gauge symmetry.\cite{RS01} 
\bea
S_{5dV}=\intxz\sqrt{-G}(-\fourth F_{MN}F^{MN}-\half m^2A^M A_M)\ ,\ 
\pr 
\label{5dEM8c}
\eea 
The 5D mass parameter $m$ is regarded as an IR-regularization
parameter. 
The 5D gauge theory is the limit $m=0$. 
For the general $m$, Casimir energy is some integral of 
 the (modified) Bessel functions with the number $\n=\sqrt{1+m^2/\om^2}$ 
where $\om$ is the warp parameter. 
The simplest case for analysis is not $m=0$ but $m=i\om$. 
UV behaviour, which is the key point of the present paper, is considered 
independent of the IR regularization parameter $m$. 
Hence, the massive case is (practically) important for the 5D theories. 
We also include the longitudinal component to respect the manifest 
5D Lorentz invariance. 
For the later use of the comparison with the warped case, we consider, instead of the 5D EM, 
the system of four 5D massless scalars with the even parity and one with the odd parity. 
          }  
The total action can be written as
\bea
\int_{-l}^l dy(\Lcal_{EM}+\Lcal_g)=\nn
2l\left\{
\half\sum_{n\in\bfZ}a_{n\m}\left(\pl^2-\left(\npl\right)^2\right)a_n^\m
+\half\sum_{n\in\bfZ,n\neq 0}b_n\left(\pl^2-\left(\npl\right)^2\right)b_n
     \right\}
.
\label{5dEM9}
\eea
Then Casimir energy $E_{Cas}$ is given by
\footnote{
Casimir energy is defined to be the free part (independent of the coupling)
of vacuum energy that depends on the {\it boundary}. The quantity is defined
to be the energy per unit space-volume of the ``brane". In the present model
of 3-brane (3+1 dim real world),																																																		 space-volume has the dimension of $L^3$. 
Hence the dimension of $E_{Cas}$ is $L^{-4}$.  
                                           }
\bea 
e^{-l^4E_{Cas}}=\int\prod_{n,\m}\Dcal a_n^\m\prod_{m\neq 0}db_m
\exp i\intxy (\Lcal_{EM}+\Lcal_g)\nn
=\exp\left[
-\half l^4\intp\left\{
4\sum_{n\in\bfZ}\ln (p^2+{m_n}^2)+\sum_{n\in\bfZ,n\neq 0}\ln (p^2+{m_n}^2)
                \right\}
      \right]
\com
\label{5dEM10}
\eea
where $p^2\equiv p_\m p^\m$ and $m_n=\npl$. This expression is the standard
one. The above KK-summation and $p_\m$ integral are divergent, hence we must
regularize them. The standard way taken by Appelquist and Chodos\cite{AC83} goes as
follows. It is sufficient to consider the even-parity quantity. 
\bea
V(l)=
\half\intp\sum_{n\in\bfZ}\ln (p^2+{m_n}^2) 
\pr
\label{5dEM11}
\eea
This is the (unregularized) Casimir energy for one scalar mode with Z$_2$-parity even. 
The first step is to introduce a reference point $l_0$.  
\bea
V(l)-V(l_0)=
\half\intp\sum_{n\in\bfZ}\ln\frac{(p^2+(\npl)^2)}{(p^2+(\nplz)^2)} 
\pr
\label{5dEM12}
\eea
This procedure makes us drop the $l$-independent 
quantity. Using the well-known formula
\bea
\sum_{n=-\infty}^\infty f_n=
\int_{-\infty}^\infty dz~f(z)+
\int_{-\infty+i\ep}^{+\infty+i\ep}dz\frac{f(z)+f(-z)}{e^{-2\pi iz}-1}
\com
\label{5dEM13}
\eea
the KK-sum in Eq.(\ref{5dEM12}) is replaced by the $z$-integral:
\bea
V(l)-V(l_0)=\nn
\half\intp\left[
\int_{-\infty}^\infty dz~\ln\frac{p^2+(\zpl)^2}{p^2+(\zplz)^2}+
\int_{-\infty+i\ep}^{+\infty+i\ep}dz
\frac{2\ln\frac{p^2+(\zpl)^2}{p^2+(\zplz)^2}}{e^{-2\pi iz}-1}
                        \right]
\pr
\label{5dEM14}
\eea
We consider the spacelike 4D momentum $p_\m$: $p^2=p_\m p^\m>0$. 
Using the second formula, 
\bea
\int_{-\infty}^\infty dz~H(z)\ln\frac{z^2+a^2}{z^2+b^2}
=2\pi\int_b^a dx~H(ix)
\com
\label{5dEM15}
\eea
the first part of (\ref{5dEM14}) is evaluated as
\bea
\half\intp
\int_{-\infty}^\infty dz~\ln\frac{(p^2+(\zpl)^2)}{(p^2+(\zplz)^2)}
=(l-l_0)\intp\sqrt{p^2}
\pr
\label{5dEM16}
\eea
This integral is quintically divergent, but turns out to be cancelled
out, as shown below. As for the second part, the integrand of the $p$-integral, 
using Eq.(\ref{5dEM15}) again, is evaluated as
\bea
\int_{-\infty}^\infty dz~\frac{\ln\frac{(p^2+(\zpl)^2)}{(p^2+(\zplz)^2)}}
                               {e^{-2\pi iz}-1}
=
2\pi\int_{\frac{l_0}{\pi}\sqrt{p^2}}^{\frac{l}{\pi}\sqrt{p^2}}
\frac{1}{e^{2\pi x}-1}dx                              \nn
=-\sqpp (l-l_0)+\ln\frac{e^{l\sqpp}-e^{-l\sqpp}}{e^{l_0\sqpp}-e^{-l_0\sqpp}}
\pr
\label{5dEM17}
\eea
We see that the first term in the above final expression, after the $p$-integral, 
cancel the quintically divergent one (\ref{5dEM16}). Hence, we finally obtain
\bea
V(l)-V(l_0)=
\intp\ln\frac{e^{l\sqpp}-e^{-l\sqpp}}{e^{l_0\sqpp}-e^{-l_0\sqpp}}\nn
=\frac{1}{8\pi^2}\frac{1}{l^4}\int_0^\infty dk k^3\{
k+\ln(1-e^{-2k})-\frac{l_0}{l}k-\ln(1-e^{-\frac{2l_0}{l}k})
                                   \},(l\sqpp\equiv k)
.
\label{5dEM18}
\eea
Using the third formula
\bea
\int_0^\infty dk~ k^3\ln (1-e^{-2k})=-\frac{3}{4}\zeta(5)
\com
\label{5dEM19}
\eea
we obtain
\bea
8\pi^2[V(l)-V(l_0)]=\nn
\left(1-\frac{l_0}{l}\right)\frac{1}{l^4}\int_0^\infty dk k^4
+\frac{1}{l^4}\left\{
-\frac{3}{4}\zeta(5)+\left(\frac{l}{l_0}\right)^4\cdot\frac{3}{4}\cdot\zeta(5)
              \right\}
\pr
\label{5dEM20}
\eea
The first term is {\it quintically} divergent. We take, as the (dimensionless)
UV {\it cutoff} of the $k$-integral, $l\La$ then 
\bea
8\pi^2[V(l)-V(l_0)]=
\frac{1}{5}l\La^5
-\frac{3}{4}\frac{\zeta(5)}{l^4}
-\left(\frac{1}{5}l_0\La^5
-\frac{3}{4}\frac{\zeta(5)}{(l_0)^4}\right)
\pr
\label{5dEM20b}
\eea
Hence, we obtain Casimir energy and Casimir force for the simple system Eq.(\ref{5dEM11}) as
\bea
8\pi^2\times V(l)=
\frac{1}{5}l\La^5
-\frac{3}{4}\frac{\zeta(5)}{l^4},\ 
F^\La_{Cas}(l)=-\frac{\pl V}{\pl l}=
\left(-\frac{1}{5}\La^5
-3\frac{\zeta(5)}{l^5}\right)\frac{1}{8\pi^2},\ \nn 
\zeta(5)=1.03693\cdots
\pr
\label{5dEM21}
\eea
The first term of $V(l)$ is {\it quintically} divergent but is simply 
proportional to $l$. 
This quantity comes from the {\it UV divergences} of 5D quantum fluctuation. 
In Casimir force 
$F^\La_{Cas}=-\frac{\pl V}{\pl l}$, that part does not depend on $l$. 
If we can find a right means of avoiding the UV divergences 
(which will be proposed later)
we may drop the (divergent) constant
contribution, and we obtain Casimir force 
for the 5D electromagnetism as
\bea
F_{Cas}=5\times \left(-3\frac{\zeta(5)}{l^5}\right)\times \frac{1}{8\pi^2}
\pr
\label{5dEM22}
\eea
The minus sign denotes the {\it attractive} force.

\section{Heat-kernel approach and position/momentum propagator\label{HK}}
We reformulate the previous section using a heat kernel in order to treat 
the problem {\it without} KK expansion. 
Note that the heat-kernel method is a complete quantization procedure for 
the free theory (quadratic theory)\cite{Schwinger51}. 
Instead of the 5D gauge fields $A^M(X)$, we introduce
partially Fourier-transformed ones $A^M_p(y)=(A^\m_p(y),B_p(y))$. 
\bea
A^\m(x,y)=\intp e^{ipx}A^\m_p(y)\q:\q \mbox{P=+}\com\nn
A^5(x,y)=\intp e^{ipx}B_p(y)\q:\q \mbox{P=}-\pr
\label{HK1}
\eea
(We do {\it not} Fourier-transform the extra space ($y$) part.) 
Then the total action is given by
\bea
S=\intxy(\Lcal_{EM}+\Lcal_g)\hspace{3cm}\nn
=\intp\int_{-l}^{l}dy \left[
\half A_{\m p}(y)(-p^2+{\pl_y}^2)A^\m_p(y)+\half B_p(y)(-p^2+{\pl_y}^2)B_p(y)
                       \right]
.
\label{HK2} 
\eea
Here we restrict the $y$-integral region to $[-l,l]$ because 
it has sufficient information and can be transformed 
(by the Fourier expansion)
to the periodic form defined in $(-\infty,+\infty)$. 
The on-shell condition is given by
\bea
(-p^2+{\pl_y}^2)A^\m_p(y)=0\com\q -l\leq y\leq l\com\q \mbox{P=}+\nn
(-p^2+{\pl_y}^2)B_p(y)=0\com\q -l\leq y\leq l\com\q \mbox{P=}-
\pr
\label{HK3}
\eea
This condition, which is {\it not} used in the following, is necessary
when we consider the S-matrix. 
Casimir energy $E_{Cas}$ is given by
\bea
e^{-l^4E_{Cas}}=\int\Dcal A_{\m p}\Dcal B_p\exp\{iS\}\hspace{3cm}\nn
=\exp l^4\intp\frac{1}{2l}\int_{-l}^{l}dy\left\{
-\frac{4}{2}\ln (p^2-{\pl_y}^2)-\half\ln (p^2-{\pl_y}^2)
                                          \right\}
\pr
\label{HK4}
\eea
Using the following formula\cite{Schwinger51}
\bea
\int_0^\infty\frac{e^{-t}-e^{-tM}}{t}dt=\ln M\com\q \mbox{det}~M>0\com\q
M\ :\ \mbox{a matrix}
\com
\label{HK5}
\eea
we can {\it formally} write 
\bea
-\ln (p^2-{\pl_y}^2)=\int_0^\infty\frac{1}{t}e^{-t(p^2-{\pl_y}^2)}dt
+\mbox{divergent constant}
\com
\label{HK6}
\eea
where the divergent constant should not depend on $p$ or $y$. 
We understand that $M$ in Eq.(\ref{HK5}) is the matrix $M_{y,y'}$ labeled by
the continuous parameters $y$ and $y'$ and that $(p^2-{\pl_y}^2)$ in Eq.(\ref{HK6})
is the differential operator acting on $|y>$, a quantum state labeled
by the position $y$.
\footnote{
$<y|$ and $|y>$ 
are introduced by Dirac\cite{Dirac39} and are called the bra and ket vectors 
respectively. It is defined by the orthonormal eigenfunctions 
of the kinetic differential operator of the system. 
When we take the orthogonality relation $<y|y'>=\delh (y-y')$, 
their physical dimensions are [|y>]=$L^{-\half}$ and [<y|]=$L^{-\half}$.   
} 
The heat kernels $H_p$ and $E_p$ are in an abstract way defined by
\bea
H_p(y,y';t)=\left.<y|e^{-(p^2-\pl_y^2)t}|y'>\right|_{P=-}\com\nn
E_p(y,y';t)=\left.<y|e^{-(p^2-\pl_y^2)t}|y'>\right|_{P=+}
\pr
\label{HK9}
\eea
Hence, we obtain the final expression of $E_{Cas}$:
\bea
e^{-l^4E_{Cas}}
=(\mbox{const})\times\nn
\exp\left[
l^4\intp\int_{0}^{\infty}\frac{dt}{t}\left\{
\frac{4}{2}\mbox{Tr}~E_p(y,y';t)+\half\mbox{Tr}~H_p(y,y';t)
                                     \right\}
                          \right]
\com
\label{HK7}
\eea
where Tr represents the integral over all $y=y'$ values.
\footnote{
For the 5D free scalar with Z$_2$-parity even, Casimir energy is given by
\bea
e^{-l^4E_{Cas}}
=(\mbox{const})\times\exp\left[
l^4\intp\int_{0}^{\infty}\frac{dt}{t}
\frac{1}{2}\mbox{Tr}~E_p(y,y';t)
                          \right]
\com
\label{HK7b}
\eea
and, for that with Z$_2$-parity odd, 
\bea
e^{-l^4E_{Cas}}
=(\mbox{const})\times\exp\left[
l^4\intp\int_{0}^{\infty}\frac{dt}{t}
\frac{1}{2}\mbox{Tr}~H_p(y,y';t)
                          \right]
\com
\label{HK7c}
\eea
          } 
\bea
\Tr E_p(y,y';t)=\int_{-l}^{l}dy~E_p(y,y;t),\ 
\Tr H_p(y,y';t)=\int_{-l}^{l}dy~H_p(y,y;t)
.
\label{HK8}
\eea

The precise definitions of $H_p$ and $E_p$, with the initial condition (\ref{HK12})
 shown below, are given by the {\it heat equations}, 
\bea
\left\{\frac{\pl}{\pl t}+p^2-\pl_y^2 \right\}H_p(y,y';t)=0\com\q \mbox{P}=-\com\nn
\left\{\frac{\pl}{\pl t}+p^2-\pl_y^2 \right\}E_p(y,y';t)=0\com\q \mbox{P}=+
\pr
\label{HK10}
\eea
The solutions are, in terms of the KK-eigen-functions, given by
\bea
H_p(y,y';t)=\frac{1}{2l}\sum_{n\in \bfZ}e^{-(k_n^2+p^2)t}
\half\{e^{-ik_n(y-y')}-e^{-ik_n(y+y')} \}\com\nn
E_p(y,y';t)=\frac{1}{2l}\sum_{n\in \bfZ}e^{-(k_n^2+p^2)t}
\half\{e^{-ik_n(y-y')}+e^{-ik_n(y+y')} \}\com\nn
k_n=\npl
\com
\label{HK11}
\eea
where we use the dimensionalities of $H_p$ and $E_p$ read from (\ref{HK7}); 
[$E_p$]=[$H_p$]=$L^{-1}$. 
\footnote{
If we ignore the dimensionality,
$$
\frac{1}{2l}\sum_{n\in \bfZ}\{e^{-(k_n^2+p^2)t}/(k_n^2+p^2)^s\}
\half\{e^{-ik_n(y-y')}\mp e^{-ik_n(y+y')} \},\ s:~\mbox{real number}
$$
are the more general solutions of Eq.(\ref{HK10}).
}
The above heat kernels satisfy the following b.c.,
\bea
\lim_{t\ra +0}H_p(y,y';t)=\frac{1}{2l}\sum_{n\in \bfZ}
\half\{e^{-ik_n(y-y')}-e^{-ik_n(y+y')} \}\nn
=
\half\{\delh(y-y')-\delh(y+y') \}\com\nn
\lim_{t\ra +0}E_p(y,y';t)=\frac{1}{2l}\sum_{n\in \bfZ}
\half\{e^{-ik_n(y-y')}+e^{-ik_n(y+y')} \}\nn
=
\half\{\delh(y-y')+\delh(y+y') \}
\com
\label{HK12}
\eea
where we have introduced 
$
\delh(y)\equiv\frac{1}{2l}\sum_{n\in \bfZ}e^{-ik_ny}
$. 
With this b.c., the heat equation (\ref{HK10}) can {\it rigorously} define $H_p$ and $E_p$. 
We here introduce the {\it position/momentum propagators} $G^{\mp}_p$ as
follows: 
\bea
G^-_p(y,y')\equiv
\int_0^\infty dt~ H_p(y,y';t)=\frac{1}{2l}\sum_{n\in \bfZ}
\frac{1}{k_n^2+p^2}\half\{e^{-ik_n(y-y')}-e^{-ik_n(y+y')} \},\nn
G^+_p(y,y')\equiv
\int_0^\infty dt~ E_p(y,y';t)=\frac{1}{2l}\sum_{n\in \bfZ}
\frac{1}{k_n^2+p^2}\half\{e^{-ik_n(y-y')}+e^{-ik_n(y+y')} \}
.\ \q
\label{HK13}
\eea
They satisfy the following differential equations of {\it propagators}:
\bea
(p^2-\pl_y^2)G^{\mp}_p(y,y')=\frac{1}{2l}\sum_{n\in \bfZ}
\half\{e^{-ik_n(y-y')}\mp e^{-ik_n(y+y')} \}\nn
\equiv
\half\{\delh(y-y')\mp\delh(y+y') \}
\com
\label{HK14}
\eea
Therefore, Casimir energy $E_{Cas}$ is, from Eqs.(\ref{HK7}) and (\ref{HK11}), given by
\bea
E_{Cas}(l)=\intp\intt 2\int_0^l dy\left\{
\half\frac{1}{2l}\sum_{n\in \bfZ}e^{-(k_n^2+p^2)t}\half\{1-e^{-2ik_ny}\}
                                       \right.   \nn
                                       \left. 
+2\frac{1}{2l}\sum_{n\in \bfZ}e^{-(k_n^2+p^2)t}\half\{1+e^{-2ik_ny}\}
                                       \right\}
\pr
\label{HK15}
\eea
This expression leads to the same treatment as that in the previous section. 

Here, we introduce the {\it generalized} P/M propagators, $I_\al$(P=$-$) 
and $J_\al$(P=+) as
\bea
I_\al (p^2;y,y')\equiv
\int_0^\infty\frac{dt}{t^\al} H_p(y,y';t)                     \nn
=\int_0^\infty\frac{dt}{t^\al}\frac{1}{2l}\sum_{n\in \bfZ}
e^{-(k_n^2+p^2)t}\half\{e^{-ik_n(y-y')}-e^{-ik_n(y+y')}\}\com\q \mbox{P=}-   \com\nn
J_\al (p^2;y,y')\equiv
\int_0^\infty\frac{dt}{t^\al} E_p(y,y';t)                     \nn
=\int_0^\infty\frac{dt}{t^\al}\frac{1}{2l}\sum_{n\in \bfZ}
e^{-(k_n^2+p^2)t}\half\{e^{-ik_n(y-y')}+e^{-ik_n(y+y')}\}\com\q \mbox{P=+}
\com
\label{HK16}
\eea
where $\al$ is an arbitrary real number. 
Then we have the following relations: 
\bea
I_0(p^2;y,y')=G_p^-(y,y')\com\q J_0(p^2;y,y')=G_p^+(y,y')\com\nn
\frac{\pl I_\al(p^2;y,y')}{\pl p^2}=-I_{\al-1}(p^2;y,y'),\ 
\int_{p^2}^\infty dk^2 I_\al (k^2;y,y')=I_{\al+1}(p^2;y,y'),\nn
\frac{\pl J_\al(p^2;y,y')}{\pl p^2}=-J_{\al-1}(p^2;y,y'),\ 
\int_{p^2}^\infty dk^2 J_\al (k^2;y,y')=J_{\al+1}(p^2;y,y'),\nn
(p^2-\pl_y^2)I_\be(p^2;y,y')=-\be I_{\be+1}(p^2;y,y')\com\nn  
(p^2-\pl_y^2)J_\be(p^2;y,y')=-\be J_{\be+1}(p^2;y,y')\com
\be\neq 0
\pr
\label{HK17}
\eea
Finally, we obtain the following useful expression of Casimir energy
in terms of P/M propagators,
\bea
E_{Cas}(l)
=\intp\left\{
\frac{1}{2}\mbox{Tr}~I_1(p^2;y,y')+\frac{4}{2}\mbox{Tr}~J_1(p^2;y,y')
       \right\}                                             \nn
=\intp\int_{p^2}^\infty\left\{\half\Tr I_0(k^2;y,y')+2\Tr J_0(k^2;y,y') \right\}dk^2\nn
=\intp\int_{p^2}^\infty\left\{\half\Tr G_k^-(y,y')+2\Tr G_k^+(y,y') \right\}dk^2
\pr
\label{HK18}
\eea

Here, we list the dimensions of the various quantities appeared above.
\[
  \begin{array}{c|c|c|c|c|c|c|c|c}
\mbox{Dim} & L^{1-2\al}& L^{-4} & L^{-3/2} & L^{-1} & L^{-1/2} & L & L^2 & L^{5/2}\\
\hline
       &     & E_{Cas}&          & \mu,\La,p  &          & \ep,l,y& t   &        \\
& I_\al,J_\al&         &          & H_p,E_p&          &G^\mp_p&  &        \\
   &        &         & A^\m,A^5 &         &         &     &    & A^\m_p,B_p\\
    &        &        &          &\delh(y-y')&|y>,<y|&      &    &
   \end{array}
\]
($\La,\mu,\mbox{and }\ep$ are the regularization parameters defined below.)

The P/M propagators $G_k^\mp$, which are expressed in 
Eq.(\ref{HK13}) in the form of a summation over {\it all} modes, 
can be expressed in a {\it closed} form.
(See, for example, Eq.(36) of Ref.\citen{IM0703}.)
\bea
G_k^\mp(y,y')=\pm\frac{\cosh \ktil(|y+y'|-l)\mp\cosh \ktil(|y-y'|-l)}
{4\ktil \sinh\ktil l}\com\nn
 -l\leq y\leq l\ , -l\leq y'\leq l\ ,\ 
\ktil\equiv\sqrt{k_\m k^\m}\ ,\ k_\m k^\m>0 
(\mbox{spacelike})
\com
\label{HK19}
\eea
where the plural sign means that one corresponds to the other 
in the same position.
\footnote
{ 
The expression of Eq.(\ref{HK19}) is considered in $-l\leq y\leq l$, and the 
periodicity ($y\ra y+2l$) seems lost. If, however, we {\it Fourier-expand} 
Eq.(\ref{HK19}) in the interval, the same expression as Eq.(\ref{HK13}) is obtained. 
Note that the present treatment is crucially different from the deconstruction 
approach in that {\it all} Kaluza-Klein modes are taken into account. 
The necessity of all the KK modes was stressed in the $\del(0)$-problem 
of the bulk-boundary theory.\cite{MP9712,IM05NPB}
}
By using the above results, Casimir energy is explicitly written as
\bea
E_{Cas}(l)=\intp\int_0^ldy (F^-(\ptil,y)+4F^+(\ptil,y))\com\nn
F^-(\ptil,y)\equiv \int_{p^2}^\infty dk^2 G_k^-(y,y)
=\int_\ptil^\infty d\ktil\frac{\cosh\ktil(2y-l)-\cosh\ktil l}{2\sinh(\ktil l)}\com\nn
F^+(\ptil,y)\equiv \int_{p^2}^\infty dk^2 G_k^+(y,y)
=\int_\ptil^\infty d\ktil\frac{-\cosh\ktil(2y-l)-\cosh\ktil l}{2\sinh(\ktil l)}
\pr
\label{HK20}
\eea
This is the closed expression, not the expanded one. (In relation to the degree of freedom of the system,
the integrands of $F^\mp$ are, in a later description, graphically shown in 
Figs.\ref{FintgrdMm1L100}  and \ref{FintgrdPm10L10}.)

\section{UV and IR regularization parameters and evaluation of Casimir energy
\label{UIreg}}
The integral region of Eq.(\ref{HK20}) is displayed in Fig.~1. 
In the figure, we introduce the {\it UV and IR regularization cutoffs} 
$\m\leq\ptil\leq\La$ and $\ep\leq y\leq l$. In order to suppress 
the number of artificial parameters as much as possible, we take
the relations
\footnote{
As for the numerical values of $\La$ and $l$, they depend on the chosen energy unit
( e.g. eV and J) and the physical model concerned. For example, if we consider 
the grand unified theories, $\La=10^{19}$ GeV (Planck energy) and $l^{-1}=10^3$ GeV (Tev physics)
are strong candidates. Another case is the cosmological model. Then we take 
$\La=10^{19}$ GeV and $l^{-1}=10^{-41}$ GeV ([Cosmological Size]$^{-1}$). 
$\La$ and $l$ are huge numerical numbers when applied in the real world.  
}
:
\bea
\ep=\frac{1}{\La}\com\q \m=\frac{1}{l}
\pr
\label{UIreg1}
\eea
This is the same situation as that in the {\it lattice gauge theory} 
( unit lattice size = $(1/\La)^4\times\ep=(1/\La)^5$; total lattice size =$(1/\m)^4\times l=l^5$). 
\begin{figure}
\begin{center}
\includegraphics[height=8cm]{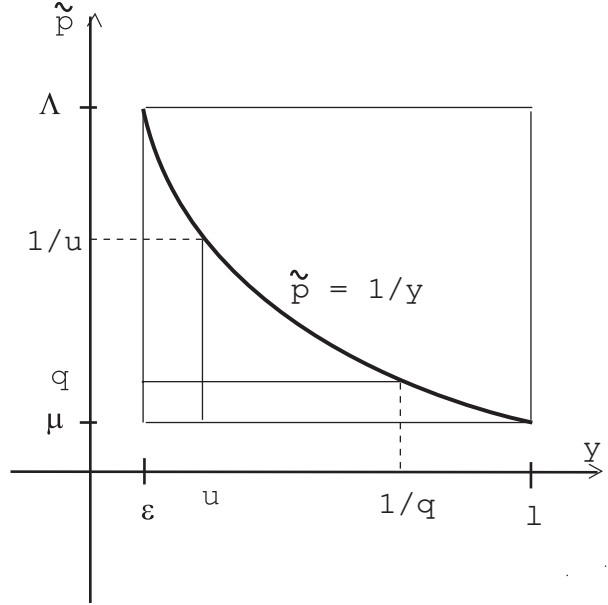}
\end{center}
\caption{
Space of (y,~$\ptil$) for integration. The hyperbolic curve 
will be used in Sec.5. 
}
\label{ypINTregion}
\end{figure}

Let us evaluate the ($\La,l$)-{\it regularized} value of Eq.(\ref{HK20}). 
\bea
E_{Cas}(\La,l)=\frac{2\pi^2}{(2\pi)^4}\int_{1/l}^{\La}d\ptil\int_{1/\La}^ldy~\ptil^3 F(\ptil,y)\com\nn
F(\ptil,y)\equiv F^-(\ptil,y)+4F^+(\ptil,y)
=\int_\ptil^\La d\ktil\frac{-3\cosh\ktil(2y-l)-5\cosh\ktil l}{2\sinh(\ktil l)}
\pr
\label{UIreg2}
\eea
The integral region of ($\ptil,y$) is the {\it rectangle} shown in Fig.~1. We now use the formula 
of the indefinite integrals:
\bea
\int\frac{\cosh x}{\sinh x}dx=\ln~\sinh(x)\com\nn
\int\frac{\cosh (ax)}{\sinh x}dx=\nn
-\frac{e^{-(1-a)x}}{1-a}\mbox{}_2F_1
                    \left(\frac{1-a}{2},1;\frac{3-a}{2};e^{-2x}\right)
 -\frac{e^{-(1+a)x}}{1+a}\mbox{}_2F_1\left(\frac{1+a}{2},1;\frac{3+a}{2};e^{-2x}\right),\nn
x>0\com\q 1>a\geq  0
\com\q
\label{UIreg3}
\eea
where $\mbox{}_2F_1(\al,\be;\ga;z)$ is Gauss's hypergeometric function. The formula below goes to the
above one in the limit $a\ra 1-0$\ (ignoring the divergent constant $1/(a-1)$). 
The integrand of $E_{Cas}(\La,l)$ (\ref{UIreg2}), $\ptil^3F(\ptil,y)$, can be {\it exactly} evaluated as
\bea
\ptil^3F(\ptil,y)=-\frac{\ptil^3}{l}\left[
-\frac{3}{2}\frac{e^{-(1-a(y))x}}{1-a(y)}\mbox{}_2F_1
                                 \left(\frac{1-a(y)}{2},1;\frac{3-a(y)}{2};e^{-2x}\right)
                                    \right.         \nn
                                    \left.
-\frac{3}{2}\frac{e^{-(1+a(y))x}}{1+a(y)}\mbox{}_2F_1
                                 \left(\frac{1+a(y)}{2},1;\frac{3+a(y)}{2};e^{-2x}\right)
+\frac{5}{2}\ln~\sinh(x)
                                     \right]_{x=\ptil l}^{x=\La l}
\com\nn
a(y)\equiv \left|2\frac{y}{l}-1\right|\com\q \La^{-1}\leq y<l\com\q  l^{-1}\leq\ptil\leq\La
\pr
\label{UIreg4}
\eea

\begin{figure}
\begin{center}
\includegraphics[height=8cm]{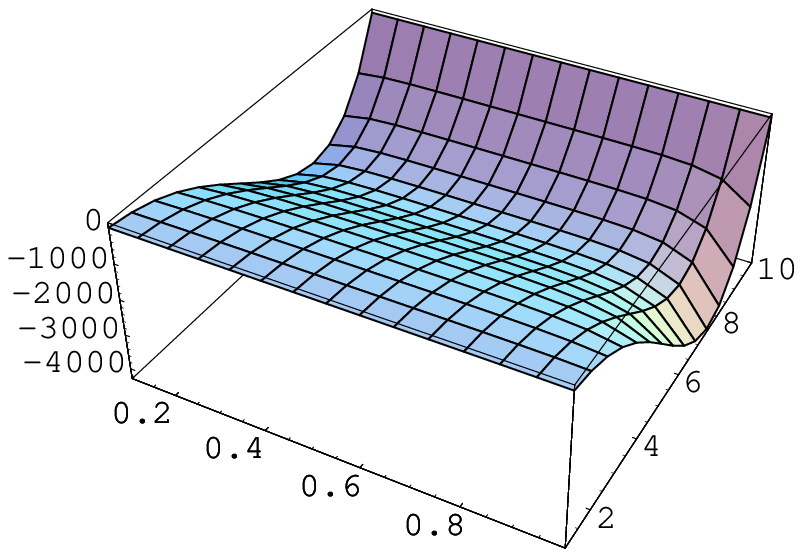}
\end{center}
\caption{
Behaviour of $\ptil^3F(\ptil,y)$ (\ref{UIreg4}). $l=1$, $\La=10$, 
$0.1\leq y<1$, $1\leq\ptil\leq 10$ . 
}
\label{p3F10La}
\end{figure}

\begin{figure}
\begin{center}
\includegraphics[height=8cm]{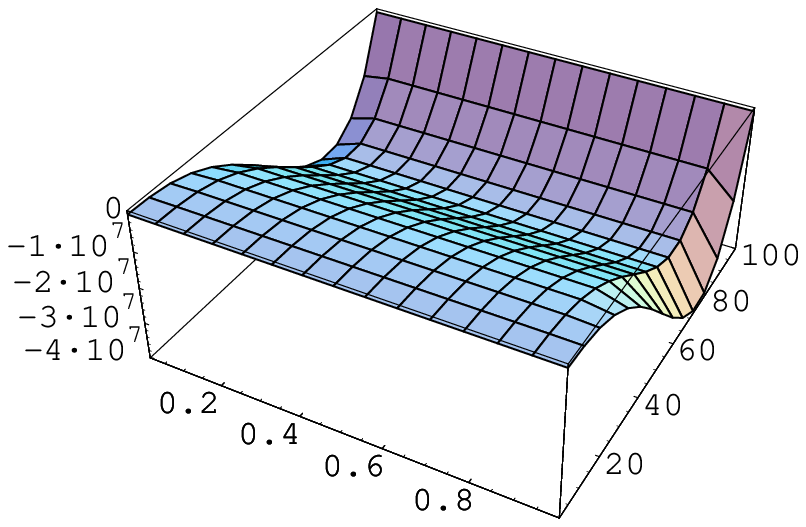}
\end{center}
\caption{
Behaviour of $\ptil^3F(\ptil,y)$ (\ref{UIreg4}). $l=1$, $\La=100$, 
$0.01\leq y<1$, $1\leq\ptil\leq 100$ . 
}
\label{p3F100La}
\end{figure}

\begin{figure}
\begin{center}
\includegraphics[height=8cm]{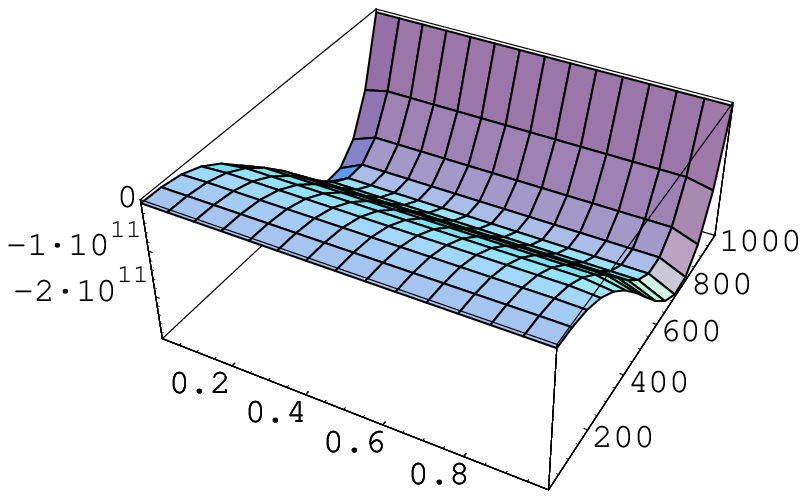}
\end{center}
\caption{
Behaviour of $\ptil^3F(\ptil,y)$ (\ref{UIreg4}). $l=1$, $\La=1000$, 
$0.001\leq y<1$, $1\leq\ptil\leq 1000$ . 
}
\label{p3F1000La}
\end{figure}

Note that Eq.(\ref{UIreg2}), with Eq.(\ref{UIreg4}), is the {\it rigorous} expression of $(\La,l)$-regularized Casimir energy. 
We show the behaviour of $\ptil^3F(\ptil,y)$ by taking the unit $l=1$ in Figs.~2--4. Three
graphs correspond to $\La=10,100,\mbox{and\ }1000$. All three graphs have a common shape. The behaviour
along $\ptil$-axis does not markedly depend on $y$. A valley `runs' parallel to the $y$-axis 
with the bottom line 
at a fixed ratio of $\ptil/\La \sim 0.75$. 
\footnote
{
The valley-bottom line (`path') $\ptil=\ptil(y)\approx 0.75\La$ corresponds to the 
solution of the minimal `action' principle:\  
$\del S=0,\ S\equiv (1/8\pi^2)\int d\ptil\int dy \ptil^3F(\ptil,y)$. This will be referred to 
in \S \ref{uncertb}. 
}
The depth of the valley is proportional to $\La^4$. 
Because $E_{Cas}$ is the ($\ptil,y$) `flat-plane' integral of $\ptil^3F(\ptil,y)$ , the {\it volume}
inside the valley is the quantity $E_{Cas}$ . It is easy to see that $E_{Cas}$ is proportional
to $\La^5$. This is consistent with Eq.(\ref{5dEM21}). 
Importantly, Eq.(\ref{UIreg2}) shows the {\it scaling} behaviour for large values of $\La$ and $l$. 
From a {\it close} numerical analysis of the ($\ptil,y$)-integral Eq.(\ref{UIreg2}), 
we confirm (see Appendix C ) that
\bea
E_{Cas}(\La,l)=\frac{2\pi^2}{(2\pi)^4}\left[ -0.1249 l\La^5
-(1.41, 0.706, 0.353)\times 10^{-5}~l\La^5\ln (l\La)\right]
\ .
\label{UIreg5}
\eea
(Note: $0.125=(1/8)\times (1/5)\times (4[even]+1[odd])$.)
This is the essentially the same result as that in \S \ref{5dEM}.
\footnote{
Note that the result of Eq.(\ref{5dEM21}) in Sec.\ref{5dEM} is obtained from the KK-expanded 
form, whereas that of Eq.(\ref{UIreg5}) in this section from the {\it closed} expression. 
The coincidence ($\La^5$ proportionality) strongly shows the correctness of both evaluations. 
At the same time, the numerical result for the first term coefficient shows that the number of significant figures
is 4. Altough the second term contibution is small, its coefficient is not stable and its significant digits 
is 1 at most. The triplet data correspond to $l=10,20$ and $40$.  
} 

\begin{figure}
\begin{center}
\includegraphics[height=8cm]{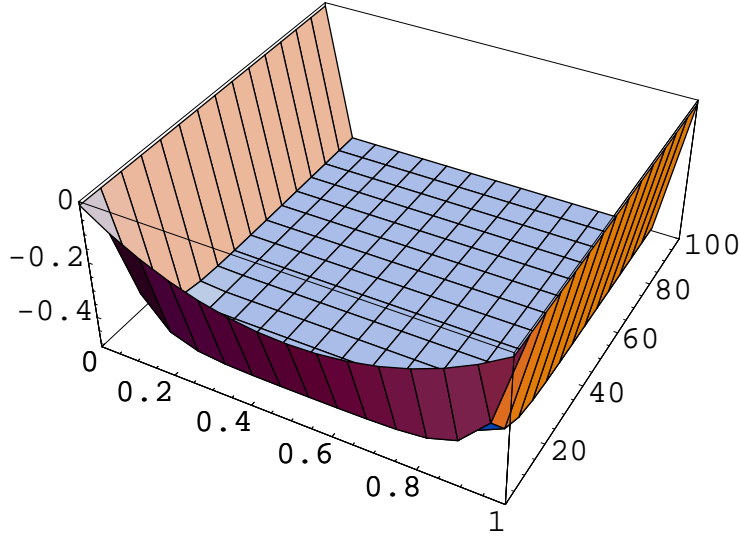}
\end{center}
\caption{
Behaviour of the integrand of $F^-$, $\mbox{Intgrd}F^-(\ktil,y;l)$ (\ref{UIreg7}). $l=1$, $\La=100$, 
$0\leq y\leq l=1$, $1\leq\ktil\leq \La=100$ . The flat plane locates at a height of $-$0.5. 
}
\label{FintgrdMm1L100}
\end{figure}

\begin{figure}
\begin{center}
\includegraphics[height=8cm]{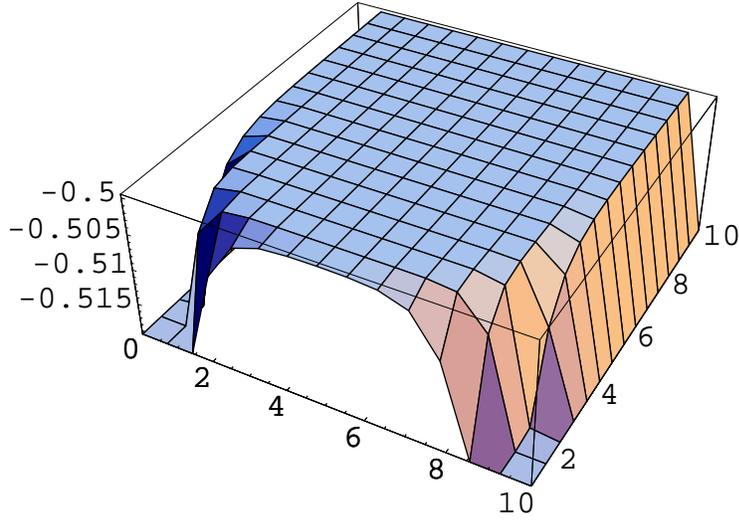}
\end{center}
\caption{
Behaviour of the integrand of $F^+$, $\mbox{Intgrd}F^+(\ktil,y;l)$ (\ref{UIreg7}). $l=10$, $\La=10$, 
$0\leq y\leq l=10$, $1\leq\ktil\leq \La=10$ . The flat plane locates at a height of $-$0.5. 
}
\label{FintgrdPm10L10}
\end{figure}

Finally we notice, from Figs.~\ref{p3F10La}--\ref{p3F1000La}, that the approximate form 
of $F(\ptil,y)$ for large values of $\La$ and $l$ is given by
\bea
F(\ptil,y)\approx -\frac{f}{2} \La \left(1-\frac{\ptil}{\La}\right)\com\q f=5
\com
\label{UIreg6}
\eea
which does {\it not} depend on $y$ or $l$.  
\footnote{
Using the approximate form Eq.(\ref{UIreg6}), $E_{Cas}$ is estimated as 
$
E_{Cas}(\La,l)\times 8\pi^2\approx
\int_{1/l}^\La d\ptil\int_{1/\La}^l dy \ptil^3(-5/2)(\La-\ptil)
= -(1/8)l\La^5(1+O(\frac{1}{\La l}))
$. This is consistent with Eq.(\ref{UIreg5}). 
}
$f$ is the {\it degree of freedom}. From this result,
we know, using the integral expression~(\ref{HK20}), 
that the following {\it approximate} form of the integrand of $F^\mp$ is 
valid in a wide range $(\ktil,y)$.  
\bea
\mbox{Intgrd}F^-(\ktil,y;l)\equiv\frac{\cosh\ktil(2y-l)-\cosh\ktil l}{2\sinh(\ktil l)}\approx -\half\com\nn
\mbox{Intgrd}F^+(\ktil,y;l)\equiv\frac{-\cosh\ktil(2y-l)-\cosh\ktil l}{2\sinh(\ktil l)}\approx -\half
\com\nn
(\ktil,y)\in \{(\ktil,y)| \ktil y\gg 1\ \mbox{and}\ \ktil (l-y)\gg 1\}
\label{UIreg7}
\eea
which can be analytically proved and is confirmed by graphically showing the above functions (see Figs.\ref{FintgrdMm1L100} and \ref{FintgrdPm10L10}. 
The table shape of the graphs 
implies the ``Rayleigh-Jeans" dominance because Casimir energy density is proportional to the 
cubic power of $\ptil$ in the region $\ptil\ll\La$. 
\footnote{
The well-known radiation spectral formula ($\be$\ is the inverse temperature):\ 
$\langle E\rangle_{\n,\be}=h\n/2~+~h\n/(e^{\be h\n}-1)$ consists of two parts. 
The first one is the zero-point energy and the second 
is Planck's ditribution part. 
The low-frequency region of Planck's dstribution formula:\ 
$h\n/(e^{\be h\n}-1)\approx 1/\be\ (\mbox{independent of }\n),\ h\n\ll 1/\be$, is called 
Rayleigh-Jeans region.  Note that the extra coordinate $y$ or $l-y$, in the present 5D model, 
plays the role of the inverse temperature $\be$. 
}

\section{UV and IR regularization surfaces, principle of 
minimal area and renormalization flow\label{surf}}

We have confirmed that heat-kernel formulation is equivalent 
to the familiar KK-expansion approach. The advantage of the
new approach is that the KK expansion is replaced by the integral 
over the extradimensional coordinate $y$ and 
all expressions are written in the closed (not expanded)
form. The $\La^5$-divergence shows notorious problem
of higher dimensional theories. 
In spite of all previous studies, 
we have not succeeded 
in defining higher-dimensional theories. (In this free theory case, 
the divergence is simple and the {\it finite} Casimir energy (force) can be read
as shown in Eq.(\ref{5dEM22}). In general, however, the divergences 
cause problems. The famous example is 
the divergent {\it cosmological constant} in the gravity-involving theories.
\cite{AC83} )
We notice here that we can solve the divergence problem if we find a way to
{\it legitimately restrict the integral region in ($\ptil,y$)-space}. 

One proposal of this was presented by Randall and Schwartz\cite{RS01}. They introduced
the {\it position-dependent cutoff},\ $\mu <\ptil <1/u\ ,~\mbox{and}~\ u\in [\ep,l]$\ , 
for the 4D-momentum integral in the ``brane" located at $y=u$. (See Fig.~\ref{ypINTregion}.)
The total integral region is the lower part of the hyperbolic curve $\ptil=1/y$. 
They succeeded in obtaining the {\it finite} $\be$-function in the 5D warped vector
model. 
We have confirmed that the $E_{Cas}$ obtained using Eq.(\ref{UIreg2}), when the Randall-Schwartz 
integral region (Fig.~\ref{ypINTregion}) is taken, is proportional to $\La^4$. 
A close numerical analysis shows (see Appendix C)
\bea
E^{RS}_{Cas}=
\frac{2\pi^2}{(2\pi)^4}\int_{1/l}^{\La}dq\int_{1/\La}^{1/q}dy~q^3 F(q,y)
=\frac{2\pi^2}{(2\pi)^4}\int_{1/\La}^{l}du\int_{1/l}^{1/u}d\ptil~\ptil^3 F(\ptil,u)\nn
=\frac{2\pi^2}{(2\pi)^4}[-8.93814\times 10^{-2}~\La^4]
\com
\label{surfM1}
\eea
which is {\it independent} of $l$. 
\footnote{
The result of Eq.(\ref{surfM1}) is consistent with the approximate form of $F$ 
obtained in Eq.(\ref{UIreg6}). 
$
(-5/2)\int_{1/l}^\La dq\int_{1/\La}^{1/q}dy q^3(\La-q)=-(1/12)\La^4(1+O(1/(\La l)^3))
$. 
$0.0893\approx 0.0833\cdots =1/12$. 
}
This shows that the divergence 
situation is indeed improved compared with the $\La^5$-divergence of Eq.(\ref{UIreg5}).   

Although Randall and Schwartz claim that holography (for the case of the warped geometry) is behind the procedure, 
the legitimateness of the restriction looks less obvious. We have proposed 
an alternative approach 
and given a legitimate explanation within the 
5D QFT\cite{IM0703,SI07Nara,SI07Osaka}. 
Here we closely examine the new regularization. 
\begin{figure}
\begin{center}
\includegraphics[height=8cm]{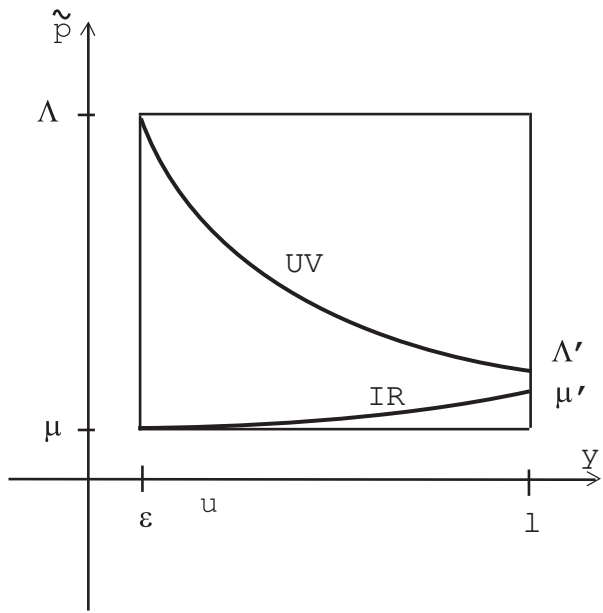}
\end{center}
\caption{
Space of ($\ptil$,y) for the integration (present proposal). 
}
\label{ypINTregion2}
\end{figure}

\begin{figure}
\begin{center}
\includegraphics[height=8cm]{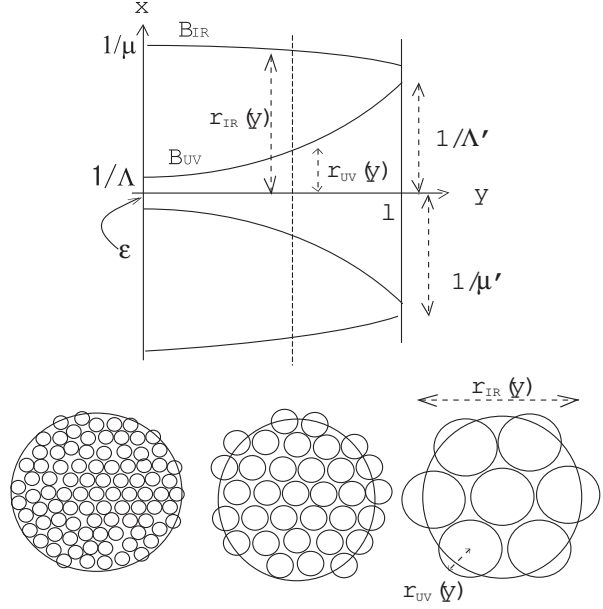}
\end{center}
\caption{
Regularization surfaces $B_{IR}$ and $B_{UV}$ in the 5D coordinate space $(x^\m,y)$, 
flow of coarse graining (renormalization) and sphere lattice regularization. 
}
\label{IRUVRegSurf}
\end{figure}

\begin{figure}
\begin{center}
\includegraphics[height=8cm]{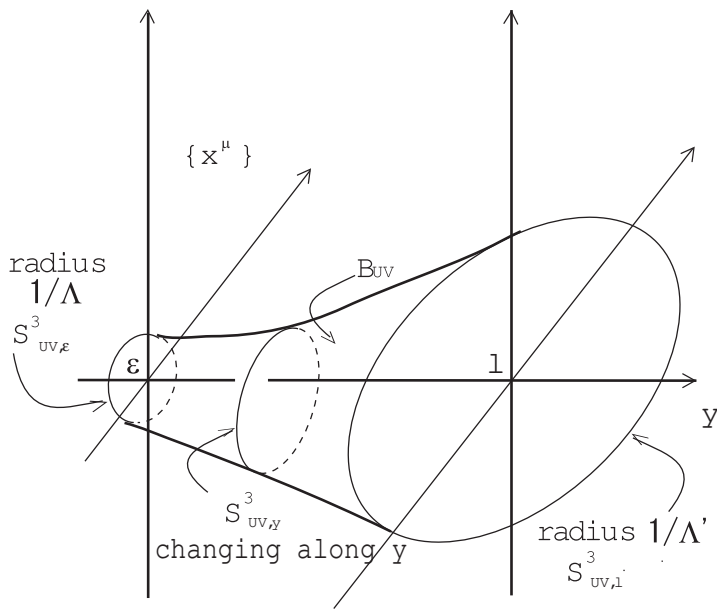}
\end{center}
\caption{
UV regularization surface in 5D coordinate space. 
}
\label{UVsurface}
\end{figure}

On the ``3-brane" at $y=\ep$, we introduce the IR cutoff $\mu$ and 
the UV cutoff $\La$\ ($\mu\ll\La$) (see Fig.\ref{ypINTregion2}), 
\bea
\mu\q \ll\q \La
\pr
\label{surf0a}
\eea
This is legitimate in the sense that we
generally perform this procedure in 4D {\it renormalizable} thoeries. 
(Here we are considering those 5D theories that are renormalizable in 3-branes. Examples are
5D electromagnetism (the present model), the 5D $\Phi^4$-theory, and the 5D Yang-Mills theory.)
For the same reason, on the
``3-brane" at $y=l$, we may have another set of IR and UV cutoffs, 
$\mu'$ and $\La'$. 
We consider the case, 
\bea
\mu'\leq\La',\ \La'\ll \La,\ \mu\sim\mu'
\pr
\label{surf0b}
\eea
This case will enable us to
introduce {\it renormalization flow}, (see our later discussion.)
We claim here that,  
as for the ``3-brane" located at each point $y$ ($\ep<y<l$), the regularization 
parameters are determined by the {\it minimal area principle}. 
\footnote{
We do {\it not} quantize the (bulk) geometry, but treat it as the {\it background}. 
The (bulk) geometry fixes the behavior of the {\it regularization} cutoff 
parameters in the field quantization. The geometry influences the ``boundary" 
of the field-quantization procedure in this way. 
}
To explain this, we move to the 5D coordinate space ($x^\m,y$), (see Fig.~\ref{IRUVRegSurf}). 
%
The $\ptil$-expression is replaced with the $\sqrt{x_\m x^\m}$-expression by 
{\it reciprocal relation}, 
\bea
\sqrt{x_\mu(y)x^\mu(y)}\equiv r(y)\q \change \q \frac{1}{\ptil(y)}
\pr
\label{surf0c}
\eea
The UV and IR cutoffs change along the $y$-axis and the trajectories form
{\it surfaces} in the 5D bulk space $(x^\mu,y)$. 
We {\it require} that the two surfaces do {\it not cross} for the purpose 
of the renormalization group interpretation (discussed later).  
We call them UV and IR regularization (or boundary) surfaces ($B_{UV}$ and $B_{IR}$), 
\bea
\mbox{B}_{UV}\q:\q\sqrt{(x^1)^2+(x^2)^2+(x^3)^2+(x^4)^2}=r_{UV}(y)\com
\q \ep=\frac{1}{\La}<y<l\com\nn
\mbox{B}_{IR}\q:\q\sqrt{(x^1)^2+(x^2)^2+(x^3)^2+(x^4)^2}=r_{IR}(y)
\com\q \ep=\frac{1}{\La}<y<l\com
\label{surf1}
\eea
The cross sections of the regularization surfaces at $y$ are the spheres $S^3$ with the 
radii $r_{UV}(y)$ and $r_{IR}(y)$. Here, we consider the Euclidean space for simplicity.
The UV-surface is stereographically shown in Fig.~\ref{UVsurface} and reminds us of the {\it closed string} propagation. 
Note that the boundary surface B$_{UV}$ (and B$_{IR}$) is the four-dimensional manifold. 

\begin{figure}
\begin{center}
\includegraphics[height=8cm]{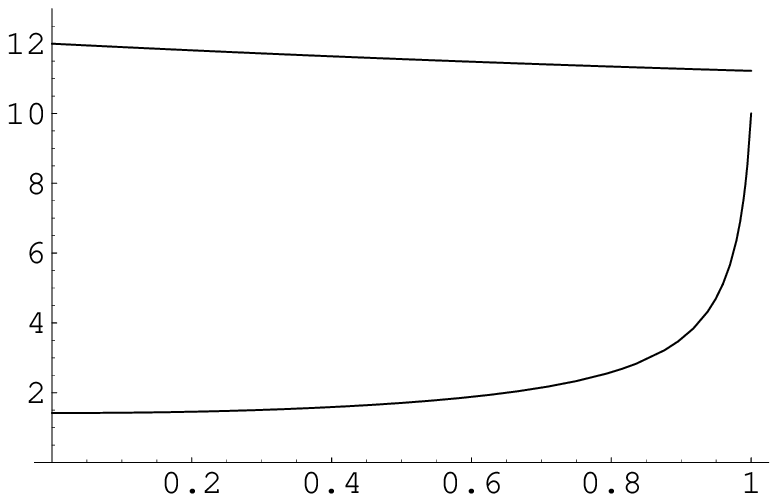}
\end{center}
\caption{
Numerical solution of Eq.(\ref{surf2}). Vertical axis: $r$; horizontal 
axis:  $0\leq y\leq l=1$. IR curve (upper): $r[0]=12.0,~ r'[0]=-1.0$\ [simply decreasing type];\ 
UV curve (lower): $r[1.0]=10.0,~ r'[1.0]=350.0$\ [simply increasing type]. 
}
\label{FlatRvsYUI12}
\end{figure}

\begin{figure}
\begin{center}
\includegraphics[height=8cm]{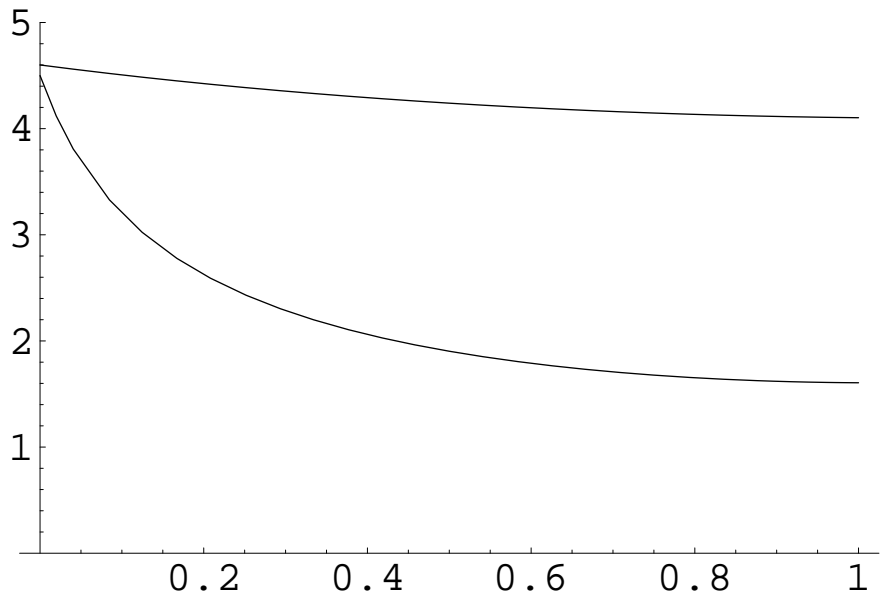}
\end{center}
\caption{
Numerical solution of (\ref{surf2}). Vertical axis: $r$; Horizontal 
axis:  $0\leq y\leq l=1$. IR curve (upper): $r[0]=4.6,~ r'[0]=-1.0$\ [simply decreasing type];\ 
UV curve (lower): $r[0]=4.5,~ r'[0]=-22.0$\ [simply decreasing type]. 
}
\label{FlatRvsYIU5}
\end{figure}

The 5D volume region bounded by $B_{UV}$ and $B_{IR}$ is the integral region 
 of Casimir energy $E_{Cas}$. 
The forms of $r_{UV}(y)$ and $r_{IR}(y)$ can be
determined by the {\it minimal area principle}, 
\bea
\del (\mbox{Surface Area})=0\com\q 
3-\frac{r\frac{d^2r}{dy^2}}{1+(\frac{dr}{dy})^2}=0\com\q 0\leq y\leq l
\pr
\label{surf2}
\eea
In Appendix A, we present the classification of all solutions (paths) and the 
general analytic solution.
In Fig.~\ref{FlatRvsYUI12}, we show two result curves of Eq.(\ref{surf2}), taking the boundary
conditions ($r'\equiv dr/dy$):
\bea
\mbox{Fig.~\ref{FlatRvsYUI12}: Fine configuration goes to coarse configuration as $y$ increases}\nn
\begin{array}{cc}
\mbox{IR curve (upper):} & r[0]=12.0, r'[0]=-1.0\q [\mbox{simply decreasing type}],\\ 
\mbox{UV curve (lower):} & r[1.0]=10.0, r'[1.0]=350.0\q [\mbox{simply-increasing type}]. 
\end{array}
\label{surf2b}
\eea
where the types of curves are specified on the basis of the classification of 
the minimal surface lines in App.~A1. 
They show that the renormalization flow shown in Fig.~\ref{ypINTregion2} really occurs
by the minimal area principle. 
In Fig.~\ref{FlatRvsYIU5}, another set of minimal surface lines are given by 
taking another boundary conditions. 
\bea
\mbox{Fig.~\ref{FlatRvsYIU5}: Coarse Conf. goes to Fine Conf. as $y$ increases}\nn
\begin{array}{cc}
\mbox{IR curve (upper):} & r[0]=4.6, r'[0]=-1.0\ [\mbox{simply decreasing type}]\com\\ 
\mbox{UV curve (lower):} & r[0]=4.5, r'[0]=-22.0\ [\mbox{simply decreasing type}]\pr 
\end{array}
\label{surf2c}
\eea
They show the opposite-direction flow of renormalization 
compared with that in Fig.~\ref{FlatRvsYUI12}. 
(See the next paragraph for the renormalization flow interpretation.) 
These two examples imply that the {\it boundary conditions} 
determine the property of the renormalization flow. 
\footnote{ 
The minimal area equation Eq.(\ref{surf2}) is the second-derivative
differential equation and  
has the general solution (see Appendix A). Hence, for the given 
two initial conditions (for example, $r(y=\ep)$ and $dr/dy|_{y=\ep}$), 
there exists a unique analytic solution. The presented graphs are 
those with these initial conditions. 
Another choice of the initial conditions, $r(y=\ep)$ and $r(y=l)$, 
is possible. 
Generally, the solution of the second-derivative differential equation 
is fixed by two {\it boundary conditions}. 
}

The present regularization scheme also gives the {\it renormalization group} interpretation
of the change in physical quantities along the extra axis (see Fig.~\ref{IRUVRegSurf}). 
\footnote{
This part is in contrast to the AdS/CFT approach where 
the renormalization flow comes from the Einstein equation of 5D supergravity. 
} 
In the ``3-brane" located at $y$, the UV cutoff is $r_{UV}(y)$
and the regularization surface is the sphere $S^3$ with the radius $r_{UV}(y)$. 
The IR cutoff is $r_{IR}(y)$
and the regularization surface is another sphere $S^3$ with a radius $r_{IR}(y)$. 
We can regard the regularization integral region as the {\it sphere lattice} of
the following properties:
\bea
\mbox{Unit lattice (cell):\ the sphere }S^3\mbox{\ with radius\ }r_{UV}(y)\ \mbox{and its inside},\nn
\mbox{Total lattice:\ the sphere }S^3\mbox{with radius\ }r_{IR}(y)\ \mbox{and its inside}.\nn
\mbox{It is composed of many cells above},\nn
\mbox{Total number of cells:\ }\mbox{const}\times\left(\frac{r_{IR}(y)}{r_{UV}(y)} \right)^4
\pr
\label{surf3}
\eea
The total number of cells changes from $(\frac{\La}{\mu})^4$ at $y=\ep$ to 
$(\frac{\La'}{\mu'})^4$ at $y=l$. Along the $y$-axis, the number increases as
\bea
\left(\frac{r_{IR}(y)}{r_{UV}(y)} \right)^4\equiv N(y)
\pr
\label{surf4}
\eea
For the ``scale" change $y\ra y+\Del y$, $N$ changes as 
\bea
\Del (\ln N)=4\frac{\pl}{\pl y}\left\{\ln \left(\frac{r_{IR}(y)}{r_{UV}(y)}\right) \right\}\cdot \Del y
\pr
\label{surf5}
\eea
When the system has some coupling $g(y)$, 
the renormalization group $\be(g)$ function (along the extra axis) is expressed as
\bea
\be=\frac{\Del(\ln g)}{\Del(\ln N)}=\frac{1}{\Del(\ln N)}\frac{\Del g}{g}
=\frac{1}{4}\frac{1}{\frac{\pl}{\pl y}\ln(\frac{r_{IR}(y)}{r_{UV}(y)})}\frac{1}{g}\frac{\pl g}{\pl y}
\com
\label{surf6}
\eea
where $g(y)$ is the renormalized coupling at $y$. 
\footnote{
Here we consider interacting theories, such as the 5D Yang-Mills theory and 
5D $\Phi^4$ theory, where the coupling $g(y)$ is the renormalized one 
in the `3-brane' at $y$.
}

We have confirmed that the minimal area principle determines the 
flow of the regularization surfaces. Among the numerical results, 
some curves are similar to the type proposed by Randall-Schwartz.   


\section{Weight function and Casimir energy evaluation\label{uncert}}
In Eq.(\ref{HK20}), Casimir energy is written by
the integral in the ($\ptil,y$) space in the range: 
$0\leq y\leq l\mbox{ and } 0\leq \ptil\leq\infty$. In \S~\ref{surf}, 
we see that {\it the integral region should be properly restricted} because the cutoff
region in the 4D world
 changes along the extra axis obeying the bulk (flat) geometry 
(minimal area principle). 
In this section, we consider an alternative version. 

\begin{figure}
\begin{center}
\includegraphics[height=8cm]{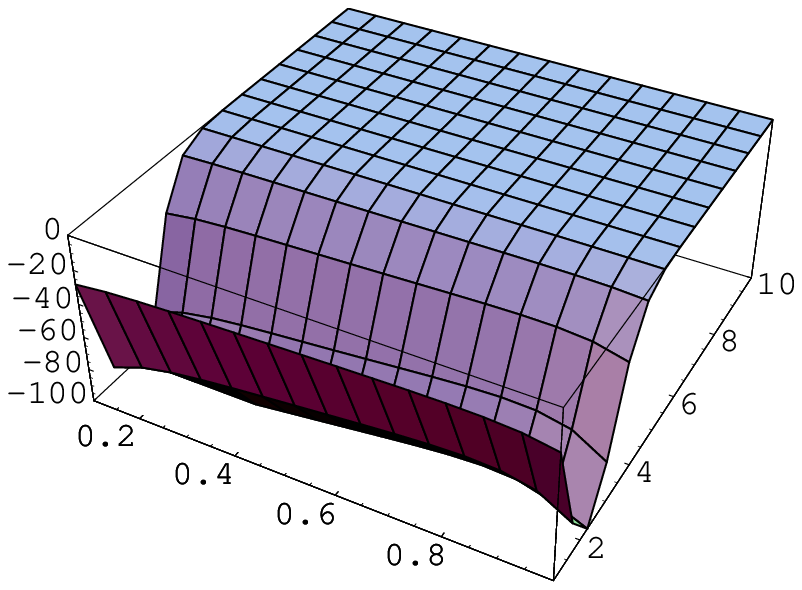}
\end{center}
\caption{
Behaviour of $\ptil^3W_1(\ptil,y)F(\ptil,y)$(elliptic suppression). 
$\La=10,\ l=1$,\  
$1/\La\leq y\leq 0.99999 l,\ 1/l\leq \ptil\leq \La$. 
}
\label{GraphW1L10m1}
\end{figure}

\begin{figure}
\begin{center}
\includegraphics[height=8cm]{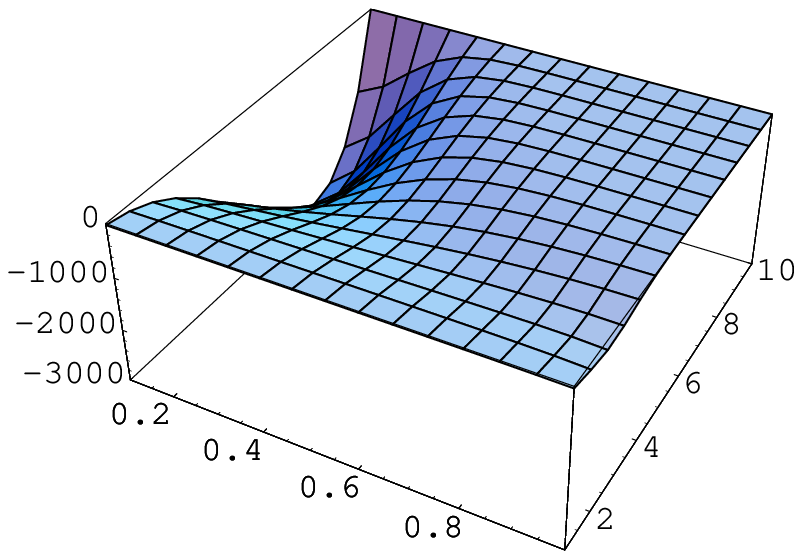}
\end{center}
\caption{
Behaviour of $\ptil^3W_3(\ptil,y)F(\ptil,y)~$(hyperbolic suppression 2). 
$\La=10,\ l=1$,\  $1/\La\leq y\leq 0.99999 l,\ 1/l\leq \ptil\leq \La$. 
}
\label{GraphW3L10m1}
\end{figure}

\begin{figure}
\begin{center}
\includegraphics[height=8cm]{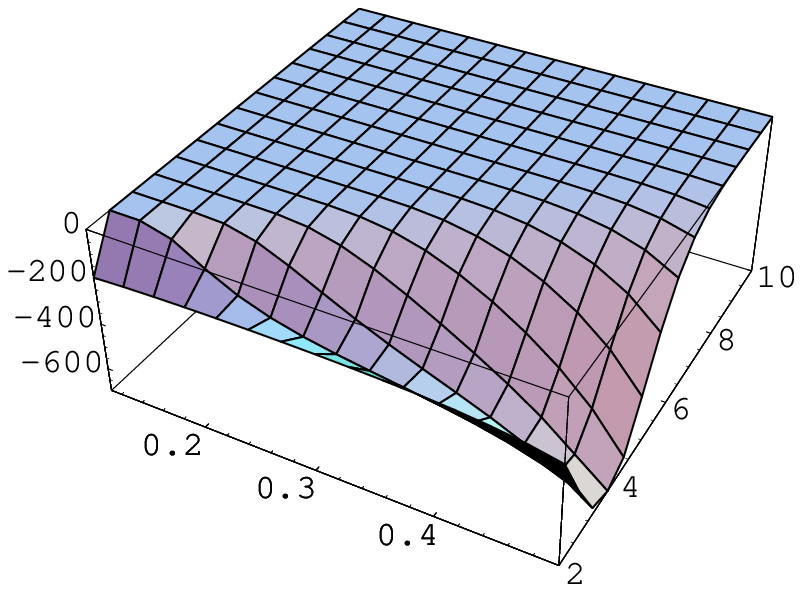}
\end{center}
\caption{
Behaviour of $\ptil^3W_4(\ptil,y)F(\ptil,y)$~(linear suppression). 
$\La=10,\ l=0.5$,\   
$1/\La\leq y\leq 0.99999 l,\ 1/l\leq \ptil\leq \La$. 
}
\label{GraphW4L10m05}
\end{figure}

\begin{figure}
\begin{center}
\includegraphics[height=8cm]{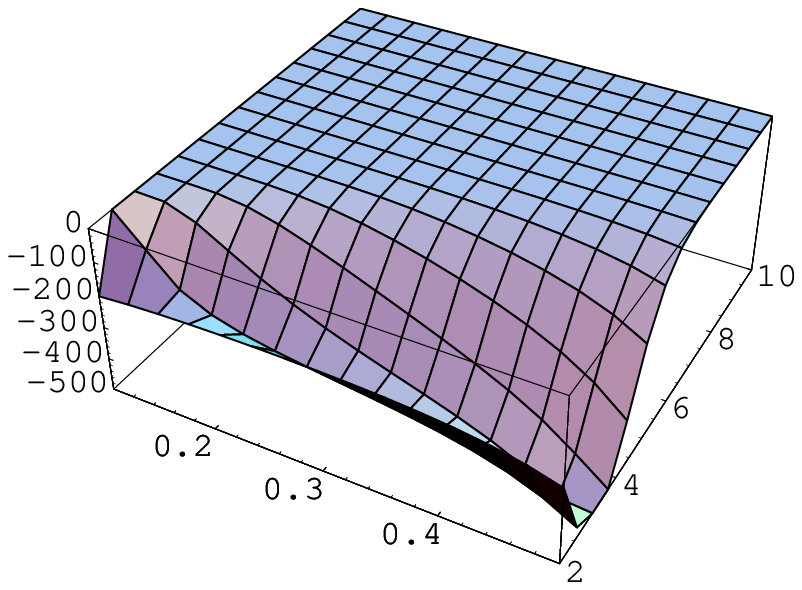}
\end{center}
\caption{
Behaviour of $\ptil^3W_6(\ptil,y)F(\ptil,y)~$(parabolic suppression 2). 
$\La=10,\ l=0.5$,\ 
$1.001/\La\leq y\leq 0.99999 l,\ 1/l\leq \ptil\leq \La$. 
The contour of this graph is given later in Fig.~\ref{ContW6La10m05}. 
}
\label{GraphW6L10m05}
\end{figure}

Instead of restricting the integral region, we introduce 
the {\it weight function} $W(\ptil,y)$ in the ($\ptil,y$)-space  
to suppress the UV and IR divergences of Casimir energy. 
\bea
E^W_{Cas}(l)\equiv\intp\int_0^ldy~ W(\ptil,y)F(\ptil,y),\ 
F(\ptil,y)\equiv F^-(\ptil,y)+4F^+(\ptil,y)\ ,\mbox{\hspace{3cm}}   \nn
\mbox{Examples of}~W(\ptil,y):\q W(\ptil,y)= \mbox{\hspace{9cm}}\nn
\left\{
\begin{array}{cc}
\frac{1}{N_1}e^{-(1/2) l^2\ptil^2-(1/2) y^2/l^2}\equiv W_1(\ptil,y),\ N_1=\frac{1.557}{8\pi^2} & \mbox{elliptic sup.},\q\\
\frac{1}{N_{1b}}e^{-(1/2) l^2\ptil^2}\equiv W_{1b}(\ptil,y),\ N_{1b}=\frac{1.820}{8\pi^2} & \mbox{kinetic-energy sup.},\q\\
\frac{1}{N_2}e^{-\ptil y}\equiv W_2(\ptil,y),\ N_2=\frac{2(l\La)^3}{8\pi^2}                   & \mbox{hyperbolic sup. 1},\q\\
\frac{1}{N_3}e^{-(1/2) \ptil^2 y^2}\equiv W_3(\ptil,y),\ N_3=\frac{2}{3}\frac{(l\La)^3}{8\pi^2}               &  \mbox{hyperbolic sup. 2},\q\\
\frac{1}{N_4}e^{-(1/2) l^4\ptil^2/y^2}\equiv W_4(\ptil,y),\ N_4=\frac{0.3222}{8\pi^2}                &  \mbox{linear sup.},\q \\
\frac{1}{N_5}e^{-l^3\ptil/y^2}\equiv W_5(\ptil,y),\ N_5=\frac{0.6342}{8\pi^2}                &  \mbox{parabolic sup. 1},\q  \\ 
\frac{1}{N_6}e^{-l^3 \ptil^2/y}\equiv W_6(\ptil,y),\ N_6=\frac{0.09788}{8\pi^2}                &  \mbox{parabolic sup. 2},\q \\  
\frac{1}{N_7}e^{-(1/2)l^4 \ptil^4}\equiv W_7(\ptil,y),\ N_7=\frac{0.3033}{8\pi^2}                &  \mbox{higher-derivative sup. 1},\q \\  
\frac{1}{N_8}e^{-(l^2/2) (\ptil^2+1/y^2)}\equiv W_8(\ptil,y),\ N_8=\frac{0.3800}{8\pi^2} & \mbox{reciprocal sup. 1},\q\\
\frac{1}{N_{47}}e^{-(l^4/2) \ptil^2(\ptil^2+1/y^2)}\equiv W_{47}(\ptil,y),\ N_{47}=\frac{0.03893}{8\pi^2} & \mbox{higher-derivative sup. 2},\q\\
\frac{1}{N_{56}}e^{-(l^3/2) (\ptil/y)(\ptil+1/y)}\equiv W_{56}(\ptil,y),\ N_{56}=\frac{0.1346}{8\pi^2} & \mbox{reciprocal sup. 2},\q\\
\frac{1}{N_{88}}e^{-(l^4/2) (\ptil^2+1/y^2)^2}\equiv W_{88}(\ptil,y),\ N_{88}=\frac{0.005006}{8\pi^2} & \mbox{higher-der reciprocal sup.},\q\\
\frac{1}{N_{9}}e^{-(l^2/2) (\ptil+1/y)^2}\equiv W_9(\ptil,y),\ N_9=\frac{0.03921}{8\pi^2} & \mbox{reciprocal sup. 3}.\q\\
\end{array}
           \right.
\label{uncert1}
\eea
where $F^\mp$ are defined in Eq.(\ref{HK20}) and $N_i$ are the normalization constants. 
\footnote
{
In the flat geometry, the periodicity parameter $l$ is a unique 
scale parameter. We make all exponents in Eq.(\ref{uncert1}) dimensionless 
using $l$. 
}
\footnote{
The normalization constants $N_i$ are defined by
\bea
\int_{l^{-1}<\ptil<\La}\frac{d^4p}{(2\pi)^4}\int_{\La^{-1}}^ldy~ W_i(\ptil,y)=
\frac{1}{8\pi^2}\frac{1}{l^3}\int_1^{l\La}dx\int_{(l\La)^{-1}}^1dw~ x^3 W_i(\frac{x}{l},lw)=\frac{1}{l^3}\com\q
l\La\gg 1\com
\label{uncert1a}
\eea
where $x\equiv l\ptil$ and $w\equiv y/l$. They are explicitly given by
\bea
8\pi^2N_1=\frac{3}{\sqrt{\e}}\int_0^1dw~ e^{-w^2/2}=1.557\com\q
8\pi^2N_{1b}=\frac{3}{\sqrt{\e}}=\int_1^\infty dx~ x^3 e^{-x^2/2}=1.820\com\nn
8\pi^2N_{2}=\int_1^\infty dx \int_{(l\La)^{-1}}^1~ dw x^3 e^{-xw}=2(l\La)^3,\ 
8\pi^2N_{3}=\int_1^\infty dx \int_{(l\La)^{-1}}^1~ dw x^3 e^{-x^2w^2/2}=(2/3)(l\La)^3,\nn
8\pi^2N_{4}=\int_1^\infty dx \int_0^1 dw~ x^3 e^{-x^2/2w^2}=0.3222\com\q
8\pi^2N_{5}=\int_1^\infty dx \int_0^1 dw~ x^3 e^{-x/w^2}=0.6342\com\nn
8\pi^2N_{6}=\int_1^\infty dx \int_0^1 dw~ x^3 e^{-x^2/w}=0.09788\com\q
8\pi^2N_{7}=\int_1^\infty dx~ x^3 e^{-x^4/2}=\frac{1}{2\sqrt{\e}}=0.3033\com\nn
8\pi^2N_8=\frac{3}{\sqrt{\e}}\int_0^1dw~ e^{-1/2w^2}=0.3800\com\q
8\pi^2N_{47}=\int_1^\infty dx \int_0^1 dw~ x^3 e^{-x^2(x^2+1/w^2)/2}=0.03893\com\nn
8\pi^2N_{56}=\int_1^\infty dx \int_0^1 dw~ x^3 e^{-(x/w)(x+1/w)/2}=0.1346\com\nn
8\pi^2N_{88}=\int_1^\infty dx \int_0^1 dw~ x^3 e^{-(1/2)(x^2+1/w^2)^2}=0.005006\com\nn
8\pi^2N_{9}=\int_1^\infty dx \int_0^1 dw~ x^3 e^{-(1/2)(x+1/w)^2}=0.03921\pr\q\q
\label{uncert1ab}
\eea
} 
Above, 
we list some examples expected for the weight function $W(\ptil,y)$. 
$W_2$ and $W_3$ are considered to correspond to the regularization taken by
Randall-Schwartz. 
How to specify the form of $W$ is the subject of the next section. 
We show the shape of the energy integrand $\ptil^3W(\ptil,y)F(\ptil,y)$ in 
Figs.~\ref{GraphW1L10m1}--\ref{GraphW6L10m05}. 
We notice that the valley-bottom line $\ptil\approx 0.75\La$, which appeared in the unweighted 
case (Figs~.2--4), is replaced by new lines:\ 
$\ptil\approx \mbox{const}$~(Fig.~\ref{GraphW1L10m1}
,$W_1$), $\ptil y\approx \mbox{const}~$(Fig.~\ref{GraphW3L10m1},$W_3$), 
$\ptil\approx \mbox{const}\times y$~(Fig.~\ref{GraphW4L10m05},$W_4$), 
$\ptil\approx \mbox{const}\times\sqrt{y}$~(Fig.~\ref{GraphW6L10m05},$W_6$). 
They are all located away from the original $\La$-effected line: $\ptil\sim 0.75\La$. 


We can check the divergence (scaling) behaviour of $E^W_{Cas}$ by 
numerically evaluating the $(\ptil,y)$-integral (\ref{uncert1}) for 
the rectangle region of Fig.~1. 
\footnote{
See Appendix C for the explanation of the numerical derivation.
} 
\bea
E^W_{Cas}=\mbox{\hspace{10cm}}\nn
\nn
\left\{
\begin{array}{cc}
-(2.500,2.501,2.501)\frac{\La}{l^3}+(-0.142,1.09,1.13)\times 10^{-4}\frac{\La\ln (l\La)}{l^3} & \mbox{for}\q W_1, \\
-(2.502,2.502,2.502)\frac{\La}{l^3}+(2.40,2.44,1.84)\times 10^{-4}\frac{\La\ln (l\La)}{l^3} & \mbox{for}\q W_{1b}, \\
-(6.0392,6.0394,6.03945)\times 10^{-2} \frac{\La}{l^3}\q\q\q\q\q                                                &     \\
                                       \q\q\q\q\q -(24.7,2.79,1.60)\times 10^{-8}\frac{\La\ln (l\La)}{l^3}&\mbox{for}\q W_2,\\
-(10.650,9.21917,9.21915)\times 10^{-2}\frac{\La}{l^3} \q\q\q\q\q                                                          & \\
\q\q\q\q\q+(153.3,1.9629624,1.9629620)\times 10^{-5}\frac{\La\ln (l\La)}{l^3}&\mbox{for}\q W_3, \\
-(2.55,2.53,2.52)\frac{\La}{l^3}+(4.56,2.63,1.48)\times 10^{-3}\frac{\La\ln (l\La)}{l^3} & \mbox{for}\q W_4,\\
-(2.55,2.54,2.52)\frac{\La}{l^3}+(6.10,3.63,2.10)\times 10^{-3}\frac{\La\ln (l\La)}{l^3} &\mbox{for}\q W_5, \\
-(2.532,2.519,2.511)\frac{\La}{l^3}+(3.19,1.83,1.03)\times 10^{-3}\frac{\La\ln (l\La)}{l^3}  & \mbox{for}\q W_6,\\
-(2.51,2.51,2.50)\frac{\La}{l^3}+(8.51,5.51,3.36)\times 10^{-4}\frac{\La\ln (l\La)}{l^3} & \mbox{for}\q W_{7}, \\
-(2.52,2.51,2.51)\frac{\La}{l^3}+(19.5,11.6,6.68)\times 10^{-4}\frac{\La\ln (l\La)}{l^3}  & \mbox{for}\q W_8,\\
-(2.55,2.55,2.55)\frac{\La}{l^3}+(5.47,5.32,5.01)\times 10^{-3}\frac{\La\ln (l\La)}{l^3}  & \mbox{for}\q W_{47},\\
-(2.54,2.53,2.52)\frac{\La}{l^3}+(4.30,2.48,1.40)\times 10^{-3}\frac{\La\ln (l\La)}{l^3}  & \mbox{for}\q W_{56},\\
-(2.61,2.61,2.60)\frac{\La}{l^3}+(1.20,1.17,1.10)\times 10^{-2}\frac{\La\ln (l\La)}{l^3}  & \mbox{for}\q W_{88},\\
-(2.54,2.52,2.51)\frac{\La}{l^3}+(3.75,2.16,1.22)\times 10^{-3}\frac{\La\ln (l\La)}{l^3}  & \mbox{for}\q W_{9}.
\end{array}
           \right.
\label{uncert1b}
\eea
The above fitting is obtained by taking the data for the range: $l=(10,20,40)$, $\La= 10 \sim 10^3$. 
The round-bracketed triplet data, corresponding to three values of $l$, 
should be the same if the scaling region is properly examined. 
The small fluctuation in the last digit number tells us the significant figures. 
The leading terms (linearly divergent terms) are firmly obtained, and the number of the significant figures (NSF) $\geq 3$. 
The log terms are obtained very poorly, and the NSF is 1 at best. 
\footnote{
No data instability appears in a warped case~\cite{SI07Osaka,SI08Singap,SI0812}. 
}
The divergent normalization constants of $W_2$ and $W_3$, namely, $N_2$ and $N_3$, are consistent with Eq.(\ref{surfM1}). 
\footnote{
In particular, $W_2$ case is more similar to that in Eq.(\ref{surfM1}) in that the log term vanishes 
in the present numerical precision. 
}
After normalizing the factor $\La l$, {\it only the log-divergence} remains. 
\bea
E^W_{Cas}/\La l =-\frac{\al}{l^4}\left( 1-4c\ln (l\La) \right) \com
\label{uncert1c}
\eea
where $\al$ can be read from Eq.(\ref{uncert1b}). 
(For the 5D EM (\ref{5dEM21}), $\al$ corresponds to the value $5\times 3\zeta(5)/(4\times 8\pi^2)\approx 3.86/8\pi^2$.)
The numerical results say that $\al$ does not so much depend on the choice of $W$ except $W_2$ and $W_3$. 
As for $c$, we expect that it reaches a fixed value as $l$ increases further. Although 
the present approach leaves the weight function $W(\ptil,y)$ unspecified, and the numerical 
results involve some ambiguity, we can say that $\al$ is approximately +2.5 (a positive number on the order of
$O(1)$), and $c$ is a positive number on the order of $O(10^{-4})$ to $O(10^{-3})$.  (The hyperbolic cases, 
$W_2$ and $W_3$, are exceptions. This is due to the divergent normalization factors $N_2$ 
and $N_3$ shown in Eq.(\ref{uncert1})).
At present, we cannot discriminate which weight is the right one. Here, we list 
the characteristic features 
(advantageous (Yes) or disadvantageous (No), independent (I) or dependent (D), and 
singular (S) or regular (R)) for each weight from the following points.
\begin{description}
\item[point 1]\ 
The behavior of $W$ for the limit $l\ra 0$ or $l\ra \infty$. This property is related to the continuity
to the ordinary field quantization.
\item[point 2]\ 
The path (bottom line of the valley) is independent (I) of the scale $l$ or dependent (D) on it. 
\item[point 3]\ 
Regular (R) or singular (S) at $y=0$.
\item[point 4]\ 
Symmetric for $l\ptil\change y/l$.
\item[point 5]\ 
Symmetric for $\ptil\change 1/y$, (Reciprocal symmetry).
\item[point 6]\ 
The Casimir energy is finite.
\item[point 7]\ 
Value of $\al$.
\item[point 8]\ 
Under Z$_2$-parity: $y\leftrightarrow -y$; $W(\ptil,y)$ is even (E), odd (O) or none (N). 
\end{description}  

\[
  \begin{array}{|c|c|c|c|c|c|c|c|c|}
\mbox{type} & W_1& W_{1b} & W_2 & W_3 & W_4 & W_5 & W_6 & W_7              \\
\hline
\mbox{point 1}      &            &                   &  &  &  &  &  &      \\
l\ra 0      &  l\del(y)          &  /              & / & / & / & / & / & / \\
l\ra \infty &  l^{-1}\times &  l^{-1}\times & / & / & l^{-2}\times & l^{-3/2}\times &
                                                   l^{-3/2}\times & l^{-2}\times   \\
            &  \del(\ptil) &  \del(\ptil) & / & / & \del(\ptil/y) & \del(\sqrt{\ptil}/y) &
                                      \del(\ptil/\sqrt{y}) &\del(\ptil^2)    \\
\mbox{point 2}  &  D &  D & I & I & D & D & D & D       \\
 \mbox{point 3} &  R &  R & R & R & S & S & S & R       \\
\mbox{point 4}  &  Y &  / & Y & Y & / & N & N & /       \\
\mbox{point 5}  &  / &  / & / & / & Y & N & N & /       \\
 \mbox{point 6} &  Y &  Y & Y & Y & Y & Y & Y & Y       \\
\mbox{point 7}  & 2.5& 2.5& 0.060& 0.092& 2.5& 2.5& 2.5& 2.5       \\
\mbox{point 8}  & E  &  E & O & E & E & E & O & E      
   \end{array} 
\]

\[
  \begin{array}{|c|c|c|c|c|c|}
\mbox{type} & W_8 & W_{47} & W_{56} & W_{88} & W_9 \\
\hline
\mbox{point 1}  &   &  &  &  &       \\
l\ra 0  &  / & / & / & / & /      \\
l\ra \infty  &  l^{-1}\times & l^{-2}\times  &
                  l^{-3/2}\times   & l^{-2}\times  & l^{-1}\times    \\
                 &  \del(\sqrt{\ptil^2+\frac{1}{y^2}}) & \del(\ptil\sqrt{\ptil^2+\frac{1}{y^2}})  &
                  \del(\sqrt{\ptil/y}\sqrt{\ptil+\frac{1}{y}})   & \del(\ptil^2+\frac{1}{y^2})  & \del(\ptil+\frac{1}{y})    \\
\mbox{point 2}  &  D & D & D & D & D       \\
 \mbox{point 3} &  S & S & S & S & S       \\
\mbox{point 4}  &  / & N & / & / & /     \\
\mbox{point 5}  &  Y & N & Y & Y & Y      \\
 \mbox{point 6} &  Y & Y & Y & Y & Y      \\
\mbox{point 7}  &  2.5& 2.6& 2.5& 2.6&2.5\\
\mbox{point 8}  &  E & E & N & E & N
   \end{array} 
\]

So far as the legitimate reason of the introduction of $W(\ptil,y)$ is not clear, 
we should regard this procedure as a {\it regularization} for 
defining higher-dimensional theories. 
We give a definition of $W(\ptil,y)$ and a legitimate explanation
in the next section. 
It should be conducted, in principle, consistently with the bulk geometry and the gauge
principle.

\section{Definition of weight function and dominant path\label{uncertb}}

In the previous section, the weight function $W(\ptil,y)$ is introduced
as some trial functions for suppressing the UV and IR divergences. 
In this section, we define (or specify) the weight function $W(\ptil,y)$ properly 
and give a {\it legitimate reason} for the introduction of $W$. 

First, the requirement for controlling the $\La^5$-divergence in \S~\ref{UIreg} 
led us to introduce some ``damping" function $W(\ptil,y)$, as in Eq.(\ref{uncert1}). 
Casimir energy is obtained by integrating out $W(\ptil,y)F(\ptil,y)$ over the entire(5D) 
space region. Among all configurations involving the integral, there exists 
a dominant configuration that contributes to the integral most dominantly. The present 
claim is that the dominant configuration should be fixed by the 5D geometry, that is, 
the 5D flat space(-time) with the periodic boundary condition and $Z_2$-symmetry. 
In the integral Eq.(\ref{uncert1}), the {\it dominant path} $\ptil=\ptil_W(y)$ 
is characterized by the differential equation obtained by the variation method: 
$\ptil\ra \ptil+\del\ptil,~y\ra y+\del y$ in the ($\ptil,y$)-integral expression Eq.(\ref{uncert1}). 
\footnote{
When $W(\ptil,y)=1$, the dominant path appears as $\ptil(y)\approx 0.75\La$~(ind. of $y$) 
in Figs.~2--4. This path, however, is ``artificially" created by the UV cutoff. It is 
irrelevant to the 5D geometry. The differential equation of $\ptil_W(y)$ is obtained 
in Eq.(\ref{WeiGeo1}). 
}
Here, we {\it require}, on the basis of the present claim, 
that the dominant path $\ptil_W(y)$ coincides with the curve $r=r_g(y)$ 
(or its momentum counterpart $\ptil_g(y)$, (\ref{rcp4})) 
which is determined by the minimal area condition Eq.(\ref{surf2}). 
Note that the minimal area curve $r=r_g(y)$ is defined by the 5D geometry.      
\footnote{
In \S~5, we required the minimal area condition on the UV and IR 
regularization surfaces (boundary configuration),
whereas, in this section, we require it on the dominant configuration 
in the $(\ptil,y)$ or $(r,y)$-integral appeared in the expression of Eq.(\ref{uncert1}). 
In other words, we have {\it fixed} the `dominant' configuration (path), 
around which a small (``quantum") fluctuation may occur, 
by taking the 
minimal surface curve. 
}

To explain the previous paragraph using only the coordinate ($r=\sqrt{x^ax^a},y$), we move to the coordinate expression by 
partial Fourier transformation.  
Casimir energy in Eq.(\ref{uncert1}) is re-expressed as
\bea
\Fhat(r(x),y)
=\intp e^{ipx}F(\ptil,y)\com\q
\What(r(x),y)
=\intp e^{ipx}W(\ptil,y)\com\nn
r(x)\equiv\sqrt{(x^1)^2+(x^2)^2+(x^3)^2+(x^4)^2}\com\nn
E^W_{Cas}(l)=\intfx\int_0^ldy~ \What(r(x),y)\Fhat(r(x),y)=\nn
2\pi^2 \int_0^ldy \int_0^\infty dr\exp\{
3\ln r+\ln\What(r,y)+\ln\Fhat(r,y)
                                     \}
\pr
\label{uncert2}
\eea
The unweighted case is $\What(r(x),y)=\del^4(x), W(\ptil,y)=1$. 
The dominant contribution (path) $r=r_W(y)$ to $E^W_{Cas}$ is given by 
the {\it minimal `action' principle}, that is, by applying  
the steepest-descend method to Eq.(\ref{uncert2}). 
\bea
\frac{d r}{d y}=
\frac{   -\frac{1}{\What}\frac{\pl\What}{\pl y}
                                 -\frac{1}{\Fhat}\frac{\pl\Fhat}{\pl y}   }
     {    \frac{3}{r} +\frac{1}{\What}\frac{\pl\What}{\pl r}
                                   +\frac{1}{\Fhat}\frac{\pl\Fhat}{\pl r} }
\equiv {\hat \Vcal}_1(\What,\pl_r\What,\pl_y\What;r,y)
\pr
\label{uncert3}
\eea
(The valley-bottom lines that appeared in Figs.~\ref{GraphW1L10m1}--\ref{GraphW6L10m05} 
are regarded as the dominant paths.) 
Using the above result, we can obtain $d^2r/dy^2$. 
\bea
{\hat \Wcal}_y\equiv   \frac{1}{\What}\frac{\pl\What}{\pl y}
                                 +\frac{1}{\Fhat}\frac{\pl\Fhat}{\pl y} \com\q
{\hat \Wcal}_r\equiv    \frac{1}{\What}\frac{\pl\What}{\pl r}
                                   +\frac{1}{\Fhat}\frac{\pl\Fhat}{\pl r} \com\nn
\frac{d^2r}{dy^2}=-\frac{\pl_y{\hat \Wcal}_y}{3r^{-1}+{\hat \Wcal}_r}
+\frac{{\hat \Wcal}_y (\pl_r{\hat \Wcal}_y+\pl_y{\hat \Wcal}_r)}{(3r^{-1}+{\hat \Wcal}_r)^2} 
-\frac{{\hat \Wcal}_y^2 (-3r^{-2}+\pl_r{\hat \Wcal}_r)}{(3r^{-1}+{\hat \Wcal}_r)^3}\nn
\equiv{\hat \Vcal}_2(\What,\pl_r \What,\pl_y \What,\pl^2_r \What,\pl_y\pl_r \What,
\pl^2_y \What; r,y)
\pr
\label{uncert3b}
\eea
We {\it require} here that the path $r=r_W(y)$ of (\ref{uncert3}), which is 
defined in a $\What (r,y)$-dependent way, 
coincides with the minimal surface curve $r_g(y)$ (\ref{surf2}), 
which is defined independently of $\What(r,y)$. Hence, $\What(r,y)$ 
is defined by inserting Eqs.(\ref{uncert3}) and (\ref{uncert3b}) in Eq.(\ref{surf2}):  
\bea
{\hat \Vcal}_2(\What,\pl_r \What,\pl_y \What,\pl^2_r \What,\pl_y\pl_r \What,
\pl^2_y \What; r,y)   \nn
-\frac{3}{r}\{{\hat \Vcal}_1(\What,\pl_r\What,\pl_y\What;r,y)\}^2-\frac{3}{r}=0
\pr
\label{uncert3c}
\eea
We call this equation the ``W-defining equation". 
It defines the weight function $\What(r,y)$
in terms of the bulk metric (geometry) and model information $\Fhat$. 
(Note: $\Fhat$ is given by Eq.(\ref{UIreg2}).) 
In Appendix B.2, we treat the W-defining equation in ($\ptil=\sqrt{p^ap^a},y$) variables. 
Besides, how much the trial weight functions satisfy 
the above definition is numerically examined. 

The scaling of the renormalized coupling $g(y)$ is given by
\bea
\be
=-\frac{1}{4}\frac{1}{\frac{\pl}{\pl y}\ln r(y)}\frac{1}{g}\frac{\pl g}{\pl y}
\com
\label{uncert4}
\eea
where $g(y)$ is a renormalized coupling at $y$.  
In the above derivation, Eq.(\ref{surf6}) is 
used in the case\ $\frac{\pl}{\pl y}r_{IR}(y)=0,\mbox{ and }r_{UV}(y)=r(y)$.
 
\section{Discussion and conclusion\label{conc}}

Let us suppose that we have found the right weight function and that 
the divergences are successfully suppressed logarithmically.
Casimir energy (\ref{5dEM21}) is replaced by
\bea
8\pi^2E_{Cas}=-\frac{3}{4}\frac{\zeta(5)}{l^4}(1-4c \ln(l\La))=-\frac{3}{4}\frac{\zeta(5)}{{l}'^4}
\com
\label{conc1}
\eea
where $c$ is some constant (see Eq.(\ref{uncert1c})). This shows 
that the periodicity parameter (or the compactification size) $l$  changes 
as the renormalization scale changes. 
{\it The parameter $l$ suffers from the renormalization effect}. 
It shows the field's {\it interaction with the boundaries}. 
The above relation is {\it exact} because 
the present system is the {\it free} theory, and the heat-kernel (1-loop) approach 
can be regarded as 
a {\it complete (nonperturbative)} quantum treatment\cite{Schwinger51}.  
Note that, in familiar 4D renormalizable interacting theories, 
the 1-loop effect is proportional to (coupling)$^2$. In the present case, however, 
$c$ is a {\it pure number}. 
When $c$ is regarded small, i.e., $c\ll 1$, we can approximate $l'$ as
\footnote{
From the results of Eq.(\ref{uncert1b}), $c\sim O(10^{-3})$. 
          }
\bea
l'\approx l(1+c\ln(l\La))
\com
\label{conc1b}
\eea
The scaling behaviour of $l$ is given by
\bea
\be_l~(\be\mbox{-function})=\frac{\pl}{\pl (\ln \La)}\ln\frac{l'}{l}=c
\pr
\label{conc2}
\eea
When $c>0$, the compactification size $l$ {\it grows} ({\it shrinks}) as the cutoff scale $\La$
{\it increases} ({\it decreases}), whereas when $c<0$, $l$ {\it shrinks} ({\it grows}) as the cutoff scale $\La$
{\it increases} ({\it decreases}). The former case is expected. 
When $c=0$, it means that the size $l$ has no quantum effect. 
\footnote{
In the unified theories based on the string theory, $l$ is called the {\it moduli parameter} and 
is given by the vacuum expectation value of the dilaton field. How to fix the parameter 
is the moduli stabilization problem. 
}

The present 5D geometry is flat. The interesting application is the warped case. 
The analysis is under way. Partial interesting results are obtained\cite{SI07Osaka,SI08Singap,SI0812}. 
More or less, the arguments go similarly to that in the flat case 
except that the periodic and hyperbolic functions are replaced by 
the Bessel and modified Bessel ones.  
The essential difference is that one additional massive parameter, the 5D AdS curvature, 
appears in the expressions. $E_{Cas}$ is expressed by the massive parameter in addition to
$l$ and $\La$. 

In the present standpoint, the space-time geometric field $G_{MN}$ 
is regarded as a background one. It is {\it not} quantized.  
As for other bulk fields, we assume that they are renormalizable in the 3-brane. 
The role of the geometry appears when it is {\it required} that the dominant `path', determined 
by the (EM) field quantization, coincides with the geometrically determined `path' 
({\it minimal area principle}). 
Practically $W$ plays the role of suppressing the integral by weighting 
the original integrand. 
Although we have already stated the definition of $W$ in \S~\ref{uncertb} and Appendix B, 
it is important to know the true meaning of $W$. 
In the next paragraph, we argue one possible interpretation. 

In order to most naturally accomplish 
the above requirement, we can go to a new step. That is, 
we {\it propose} to {\it replace} the 5D space integral in Eq.(\ref{uncert2}) 
with the following {\it path integral}. Namely, we 
{\it newly define} Casimir energy in the higher-dimensional theory as follows:  
\bea
\Ecal_{Cas}(l,\La)\equiv \mbox{\hspace{5cm}}\nn
\int_{1/\La}^{l}d\rho\int_{\ptil(0)=\ptil(l)=1/\rho}
\prod_{a,y}\Dcal p^a(y)F(\ptil,y)
\mbox{\ }\exp\left[ 
-\frac{1}{2\al'}\int_{0}^{l}\frac{1}{\ptil^3}\sqrt{\frac{\ptil^{'2}}{\ptil^4}+1}~ dy
    \right]\nn
=\int_{1/\La}^{l}d\rho\int_{r(0)=r(l)=\rho}
\prod_{a,y}\Dcal x^a(y)F(\frac{1}{r},y)
\mbox{\ }\exp\left[ 
-\frac{1}{2\al'}\int_{0}^{l}\sqrt{{r'}^2+1}~r^3 dy
    \right]\com
\label{conc3x}
\eea 
where the limit $\La l\ra \infty$ is taken and the surface (string) tension parameter $1/2\al'$ is introduced
(note: the dimension of $\al'$ is [length]$^4$). 
$F(\ptil,y)$ or $F(1/r,y)$ is the energy density operator induced from the quantization 
of 5D EM fields in Eq.(\ref{HK20}). 
The weight factor comes from the {\it area} suppression: \nl 
$\exp(-\mbox{Area}/2\al')=\exp[-(1/2\al')\int\sqrt{\mbox{det}g_{ab}}d^4x]$. 
In the above expression, we have followed 
the path-integral formulation of the density matrix (see Feynman's text\cite{Fey72}). 
The above definition clearly shows that the 4D space coordinates $x^a$ 
or the 4D momentum coordinates $p^a$ are {\it statistically quantized} with 
the Euclidean time $y$ and the ``{\it area} Hamiltonian" 
$H=\int\sqrt{\det g_{ab}}~d^4x$.  
\footnote{
The possibility of the quantum feature of the 5D coordinate/momenta was pointed out 
in Ref.~\cite{IM05NPB} where the idea of the phase space $(y,\ptil)$ was presented in relation 
to the divergence problem of the ``deformed" propagator in the 5D bulk-boundary theory. 
}
This reminds us of the {\it space-time uncertainty principle} \cite{Yoneya87} introduced 
in the development of the string theory. 
The 5D quantum field theory leads to some {\it quantum statistical} system of 
the 4D coordinates \{$x^a(y)$\} with the inverse temperature parameter $y$. 
In this view, 
the treatment of \S~\ref{uncert} and \ref{uncertb} is an {\it effective} action 
approach. We expect the direct evaluation of Eq.(\ref{conc3x}), numerically 
or analytically, leads to a similar result. 

In the convincing regularization procedure, there should be some limiting process. In the explanation
of the proposed one, before the previous paragraph, we do not clearly 
state the limiting process. We take large, but finite, values of $\La$ and $l$ and obtain 
some formulae through numerical calculation valid for general (large but finite) $\La$ and $l$. 
\footnote{
This situation is similar to that in the {\it lattice gauge theory}. 
}
The limiting process should be given by 
\bea
l\La\ \longrightarrow \infty
\com
\label{conc3xx}
\eea 
as in the proposed Eq.(\ref{conc3x}). In fact this limit is taken in the evaluation 
of the normalization constants of the weight functions in Eq.(\ref{uncert1}). When 
the limiting process Eq.(\ref{conc3xx}) is taken into account, the physical 
quantity, that is, Casimir energy $E_{Cas}$ (Eq.(\ref{conc1})) has the log divergence
$\ln (l\La)$. This divergence is subtracted by the renormalization procedure (\ref{conc1}) and 
(\ref{conc1b}). 
(Note that the system is a locally free theory. It does not have the mass and coupling parameters. 
The original divergences must be absorbed away by the renormalization of the wave function or 
boundary parameter(s).) The anomalous scaling behavior Eq.(\ref{conc2}) shows the non trivial 
interaction between the (free) fields and the boundaries. 

In the present analysis, 
from the beginning, the extra space is treated differently from the 4D real space(-time). 
It is regarded as the part that provides the axis for a {\it scale} change. 
The change of renormalization group is determined by the {\it minimal area condition}. 
In order to extend the quantum field theory, without the divergence problem, 
the extra space should play a role in {\it suppressing} the singular behaviour. 
Although the geometry is treated as a background, the final outcome looks to demand 
some type of quantization among the 4D coordinates (or momenta) as explained in a previous paragraph. 
The necessity of the weight 
function $W(\ptil,y)$ in Eq.(\ref{uncert1}) can be interpreted to mean that we must take the well-defined 
space(-time) measure $d^4p~dy~W(\ptil,y)$, 
instead of $d^4p~dy$ (Eq.(\ref{HK20}) or (\ref{UIreg2})), in summing over 5D space(-time). 

We have focused only on the vacuum energy. We must, of course, examine
other physical quantities such as S-matrix amplitude in the interacting theories. 

The present proposal should be compared with the string theory. 
We do not directly treat the string propagation. We start with quantizing 
the higher dimensional field theory in the standard way of QFT. We {\it require} 
the dominant configuration, the path $r=r_W(y)$, to be equal to the solution of the minimal area 
line $r_g(y)$ by introducing the weight function $W(\ptil,y)$. The bulk 
geometry takes part in the ``coordinates quantization" by {\it fixing the central configuration} 
in this way.  
The closed-string-like configuration comes into 
this formalism through the {\it minimal area principle}. 
The field theories, which are applicable to this formalism, are limited to 
those that are renormalizable on the 3-brane, such as the 5D $\Phi^4$-theory, 
5D YM theory, and 5D QED. However 
the advantages of this approach is 
that it is based on the QFT, hence we can expect various phenomenology applications. 
We stress the practical importance of calculation.

We have pointed out a possibility of quantizing 
higher dimensional field theories within QFT.

\section*{Acknowledgements}
Parts of this work have already been presented in 
the workshop ``Evolution of Space-Time-Matter in the Early Universe" 
(07.5.29, Univ. of Tokyo, Japan), ICGA8 (07.8.29-9.1, Nara Women's Univ.,~Japan), 
62nd Japan Physical Societty Meeting (07.9.21-24, Hokkaido Univ.,~Sapporo,Japan), 
KIAS-YITP Joint Workshop ``Strings, Cosmology and Phenomenology" (07.9.24-28, Kyoto Univ.,~Japan), 
the workshop ``Higher Dimensional Gauge Theory and the Unification of Forces" (07.11.29-30, Kobe Univ.,~Japan)
and the workshop ``Progress of String Theory and Quantum Field Theory"
(07.12.7-10, Osaka City Univ.). 
The author thanks T. Inagaki (Hiroshima Univ.), K. Ito (TIT), T. Kawano (Univ. of Tokyo), 
M. Peskin (SLAC) and T. Yoneya (Univ. of Tokyo) for helpful comments
and discussions on different occasions. He also thanks T. Tamaribuchi (Shizuoka Univ.) 
for advice in the computer calculation. 

\appendix
\section{Minimal surface curve in 5D flat space\label{appA}}

We analytically examine the minimal area equation (\ref{surf2}) in the 5D flat geometry. 
\bea
3-\frac{r\rddot}{1+\rdot^2}=0\com\q 0\leq y\leq l
\com
\label{MiniSur9}
\eea 
where $\rdot=dr/dy$  and $\rddot=d^2r/dy^2$. 

\subsection{Classification of minimal surface curves}
Before solving the equation, we classify all solutions 
using the differential equation above. This is useful in drawing minimal surface 
curves and confirming the noncrossing of curves. 
(The requirement comes from the renormalization-group flow interpretation of the curves. 
See a few sentences before Eq.(\ref{surf1}).) 
 
In terms of $u\equiv 1/r^2$, the above one (Eq.(\ref{MiniSur9})) can be expressed as
\bea
u(y)\equiv\frac{1}{r(y)^2}=\frac{1}{x^ax^a}>0\q;\q 
\q{\ddot u}=-6u^2\leq 0\com\q 0\leq y\leq l
\pr
\label{MiniSur10}
\eea 
From this equation, we can derive an important {\it inequality relation}:
\bea
{\dot u}|_{y=l}-{\dot u}|_{y=0}=-6\int_0^lu^2 dy<0
\pr
\label{MiniSur11}
\eea 
The inequality ${\ddot u}\leq 0$ in Eq.(\ref{MiniSur10}) implies 
that $u(y)$ is convex upwards.  

\begin{figure}
\begin{center}
\includegraphics[height=4cm]{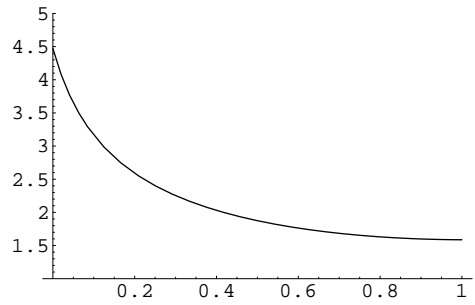}
\end{center}
\caption{
Geodesic curve $r(y)$ of Eq.(\ref{MiniSur9}) by Runge-Kutta method. Type (ia) simply decreasing. $r(0)=4.472,~\rdot(0)=-22.36$ 
}
\label{5DFlatRvsY3}
\end{figure}

\begin{figure}
\begin{center}
\includegraphics[height=4cm]{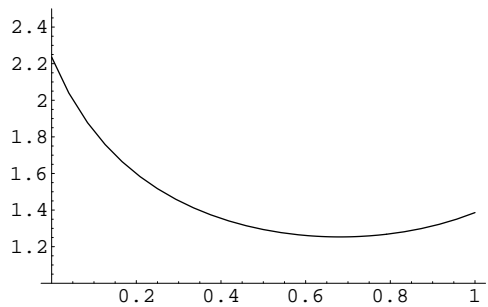}
\end{center}
\caption{
Geodesic curve $r(y)$ of Eq.(\ref{MiniSur9}) by Runge-Kutta method. 
Type (ib$\al$). $r(0)=2.236,~\rdot(0)=-5.590$. 
}
\label{5DFlatRvsY2}
\end{figure}

\begin{figure}
\begin{center}
\includegraphics[height=4cm]{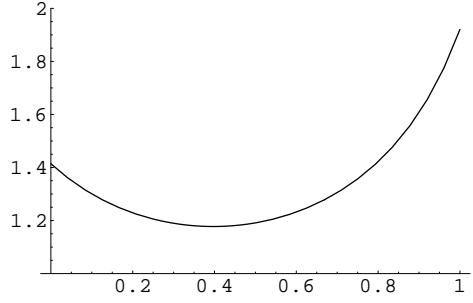}
\end{center}
\caption{
Geodesic curve $r(y)$ of Eq.(\ref{MiniSur9}) by Runge-Kutta method. 
Type (ib$\be$). $r(0)=1.4142,~\rdot(0)=-1.4142$. 
}
\label{5DFlatRvsY1}
\end{figure}

\begin{figure}
\begin{center}
\includegraphics[height=4cm]{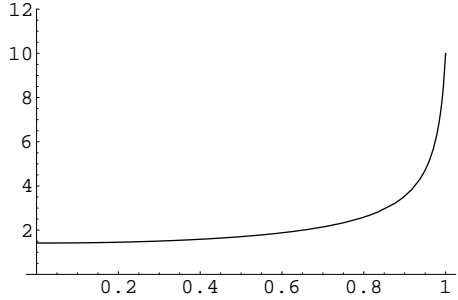}
\end{center}
\caption{
Geodesic curve $r(y)$ of Eq.(\ref{MiniSur9}) by Runge-Kutta method. 
Type (ii) simply increasing. 
$r(1.0)=10.0,~\rdot(1.0)=350.0$. 
}
\label{5DFlatRvsY7}
\end{figure}

Making use of the above relation, we can classify all solutions (paths) 
as follows. \nl
(i)\ $\udot(y=0)>0$\nl
\hspace*{20pt} (ia)\ $\udot(l)>0$\nl
\hspace*{40pt}    \parbox{410pt}{In this cae $\udot(y)>0 \mbox{\ for\ } 0\leq y\leq l$. 
$u(y)$ is simply increasing ($r(y)$ is simply decreasing),}\nl
\hspace*{40pt}    
Fig.\ref{5DFlatRvsY3}\nl
\hspace*{20pt} (ib)\ $\udot(l)<0$\nl
\hspace*{40pt}    (ib$\al$)\ $u(0)<u(l)$\ 
Fig.~\ref{5DFlatRvsY2}\nl 
\hspace*{40pt}    (ib$\be$)\ $u(0)>u(l)$\ 
Fig.~\ref{5DFlatRvsY1}\nl 
(ii)\ $\udot(y=0)<0$\nl
\hspace*{20pt} $u(y)$ is simply decreasing ($r(y)$ is simply increasing), 
Fig.~\ref{5DFlatRvsY7}\nl

Although numerical solutions are displayed in Figs.~\ref{5DFlatRvsY3}--\ref{5DFlatRvsY7}, 
the general analytic solution is given in Appendix A.2. 
We have confirmed the high-precision equality between the numerical curves
and the analytical ones.

\subsection{General analytic solution of minimal surface curve 
(Eq.(\ref{surf2}),~(\ref{MiniSur9}), or (\ref{MiniSur10}))}

We solve the differential equation (Eq(ref{surf2}),~(\ref{MiniSur9}), or (\ref{MiniSur10})), which is the
minimal surface trajectory of the 5D {\it flat} geometry. 
The first integral is given by
\bea
\frac{d}{dy}\left(\half {\udot}^2+2u^3\right)=0\com\nn
\half {\udot}^2+2u^3=2C\com\q C(>0):\ \mbox{an integral constant}\nn
\udot=\frac{du}{dy}=\pm 2\sqrt{C-u^3}\com\q 0<u\leq \sqrt[3]{C}
\pr
\label{A1}
\eea 
We see that the present classical system is equivalent to one particle mechanics 
with the potential $V=2u^3$. 
The second integral is obtained as follows: 
\bea
\int\frac{du}{\sqrt{C-u^3}}=\pm 2\int dy=\pm 2y+C',\ 
C':\mbox{another integral constant}
\ ,
\label{A2}
\eea 
The LHS of the above equation can be integrated as
\bea
v\equiv \frac{u}{\sqrt[3]{C}}\com\q 0<v\leq 1\com\nn
\int\frac{du}{\sqrt{C-u^3}}=\frac{\sqrt[3]{C}}{\sqrt{C}}\int\frac{dv}{\sqrt{1-v^3}}
=-\frac{C^{-1/6}}{\sqrt[4]{3}}
F\left(\cos^{-1}\frac{\sqrt{3}-1+v}{\sqrt{3}+1-v}, \frac{\sqrt{3}+1}{2\sqrt{2}}\right)
,
\label{A3}
\eea 
where $F(\vp,k)$ is the {\it elliptic integral of the first kind} and 
the following integral formula is used. 
\bea
\int_v^1\frac{dx}{\sqrt{1-x^3}}
=\frac{1}{\sqrt[4]{3}}
F\left(\cos^{-1}\frac{\sqrt{3}-1+v}{\sqrt{3}+1-v}, \frac{\sqrt{3}+1}{2\sqrt{2}}\right)\com\q
0<\frac{\sqrt{3}-1+v}{\sqrt{3}+1-v}\leq 1\com    \nn
F(\vp,k)=\int_0^\vp\frac{d\sh}{\sqrt{1-k^2\sin^2\sh}}
=\int_0^{z_1}\frac{dz}{\sqrt{(1-z^2)(1-k^2z^2)}}\equiv \Ftil(z_1,k)\com\nn
0\leq k\leq 1\com\q -\frac{\pi}{2}<\vp<\frac{\pi}{2}\com\q
z_1\equiv \sin\vp=\mbox{sn}(\Ftil,k)
\com
\label{A4}
\eea 
where $\mbox{sn}(\Ftil,k)$ is defined by the inverse function of $\Ftil(z_1,k)$ 
and is called {\it Jacobi's elliptic function}. It has the period $4K(k)\equiv 4F(\frac{\pi}{2},k)$. 
Hence,
\bea
z_1=\sin\vp=\pm\sqrt{1-\cos^2\vp}=\pm\{(1-\cos\vp)(1+\cos\vp)  \}^{1/2}\nn
=\pm\frac{2\sqrt[4]{3}\sqrt{1-v}}{\sqrt{3}+1-v}
=\mbox{sn}\left(-\sqrt[4]{3}C^{1/6}(\pm 2y+C'),\frac{\sqrt{3}+1}{2\sqrt{2}}\right)
\pr
\label{A5}
\eea 
We can solve the above equation w.r.t. $v=\frac{u}{\sqrt[3]{C}}$.
\footnote{
Note that one of two solutions (of a quadratic equation w.r.t. $v$) is omitted 
here because of 
$(\sqrt{3}+1)\snbar^2-2\sqrt{3}(1+|\cnbar|)=-\{(\sqrt{3}+1)|\cnbar|+\sqrt{3}-1\}
\{|\cnbar|+1\}<0$.
} 
\bea
u_+=\frac{1}{{r_+}^2}=\sqrt[3]{C}\frac{(\sqrt{3}+1)\snbarp^2-2\sqrt{3}(1-|\cnbarp|)}
{\snbarp^2}\com\nn 
u_-=\frac{1}{{r_-}^2}=\sqrt[3]{C}\frac{(\sqrt{3}+1)\snbarm^2-2\sqrt{3}(1-|\cnbarm|)}
{\snbarm^2}\com\nn
\snbarpm(y,C,C')\equiv \mbox{sn}\left(-\sqrt[4]{3}C^{1/6}(\pm 2y+C'),\frac{\sqrt{3}+1}{2\sqrt{2}}\right)\com\nn
\cnbarpm(y,C,C')\equiv \mbox{cn}\left(-\sqrt[4]{3}C^{1/6}(\pm 2y+C'),\frac{\sqrt{3}+1}{2\sqrt{2}}\right)\pr
\label{A6}
\eea 
From the requirement $u=\frac{1}{r^2}>0$, the above result indicates that 
\footnote{
Using the relation $\snbar^2=1-\cnbar^2$, we know that 
$(\sqrt{3}+1)\snbar^2-2\sqrt{3}(1-|\cnbar|)=\{\sqrt{3}-1-(\sqrt{3}+1)|\cnbar|\}
\{|\cnbar|-1\}$.
}
\bea
\frac{\sqrt{3}-1}{\sqrt{3}+1}\leq |\cnbarpm(y,C,C')|=
\left| \mbox{cn}\left(-\sqrt[4]{3}C^{1/6}(\pm 2y+C'),\frac{\sqrt{3}+1}{2\sqrt{2}}\right) \right|
\leq 1
.
\label{A7}
\eea 
We must choose $C$ and $C'$ in such a way that the above relation is valid for 
$\mbox{}^\forall y\in (0,l)$.  
$r(y=0)$ and $\left.\frac{dr}{dy}\right|_{y=0}$ are related to $C$ and $C'$ as follows: 
\bea
r(0)=C^{-1/6}\frac{|\mbox{sn0}(C,C')|}
{\left\{(\sqrt{3}+1)\mbox{sn0}(C,C')^2-2\sqrt{3}(1-|\mbox{cn0}(C,C')|)\right\}^{1/2}}\com\nn
\mbox{sn0}(C,C')=\mbox{sn}\left(-\sqrt[4]{3}C^{1/6}C',\frac{\sqrt{3}+1}{2\sqrt{2}}\right)\com\nn
\mbox{cn0}(C,C')=\mbox{cn}\left(-\sqrt[4]{3}C^{1/6}C',\frac{\sqrt{3}+1}{2\sqrt{2}}\right)\com\nn
\left.\frac{dr}{dy}\right|_{y=0}=\mp r(0)^3 \sqrt{C-\frac{1}{r(0)^6}}\com\nn
r_+(l)=C^{-1/6}\frac{|\snbarp(l,C,C')|}
{\left\{(\sqrt{3}+1)\snbarp(l,C,C')^2-2\sqrt{3}(1-|\cnbarp(l,C,C')|)\right\}^{1/2}}\com\nn
r_-(l)=C^{-1/6}\frac{|\snbarm(l,C,C')|}
{\left\{(\sqrt{3}+1)\snbarm(l,C,C')^2-2\sqrt{3}(1-|\cnbarm(l,C,C')|)\right\}^{1/2}}
\pr
\label{A8}
\eea 
The two integration constants come from the second-derivative differential 
equation (\ref{MiniSur9}). 
In the numerical approach of the Runge-Kutta method, the constants are given by
$r(y=0)$ and $\left.\frac{dr}{dy}\right|_{y=0}$. It is important to 
check, in the simple case of the flat model, 
that the numerical solution correctly produces the analytic one. 
In the warped case, we rely only on the numerical method.

\section{Weight function and minimal surface curve\label{appB}}

In \S~{\ref{uncertb}}, we have presented a specification of the weight function $W(\ptil,y)$.  
Here, we examine its validity by numerically evaluating the consistency equation, to be given later, 
for each $W$ that appeared in Eq.(\ref{uncert1}). 

\subsection{Reciprocal space}
The 4D boundary manifold described in \S~\ref{surf} is characterized by the metric,
\bea
ds^2=\left(
\del_{ab}+\frac{x^ax^b}{(r\frac{dr}{dy})^2}
     \right)   dx^a dx^b\equiv g_{ab}(x)dx^adx^b\ ,\ r=\sqrt{x^ax^a}\ ,\nn 
(\del_{ab})=\mbox{diag}(1,1,1,1), 
A=\int\sqrt{\det g_{ab}}d^4x=\int_{1/\La}^l\sqrt{r'^2+1}~r^3 dy,r'=\frac{dr}{dy}
,
\label{rcp1}
\eea
where $\{x^a;a=1,2,3,4 \}$ are the coordinates of the 4D Euclidean space manifold. 
We introduce the reciprocal coordinates $\{p^a\}$ defined by
\bea
x^a=\frac{p^a}{p^2}\com\q p^2\equiv p^ap^a\ ;\q
p^a=\frac{x^a}{x^2}\com\q x^2\equiv x^ax^a\ ;\q
x^2=\frac{1}{p^2}
\pr
\label{rcp2}
\eea
The metric (\ref{rcp1}) can be, in terms of these coordinates, rewritten as 
\bea
dx^a=\frac{1}{p^2}(-\del^{ab}+2Q^{ab})dp^b\com\q Q^{ab}\equiv\del^{ab}-\frac{p^ap^b}{p^2}\com\nn
Q^{ab}=Q^{ba},\ p^aQ^{ab}=0,\ Q^{ab}p^b=0,\ Q^{ab}Q^{bc}=Q^{ac},\ 
\del^{ab}Q^{ab}=3,\nn
ds^2=\frac{1}{(p^2)^2}
    \left(
\del_{ab}+p^2\frac{p^ap^b}{(\frac{d\ptil}{dy})^2}
     \right)   dp^a dp^b\equiv \gh_{ab}(p)dp^adp^b\com\q \ptil=\sqrt{p^ap^a}
\pr\q
\label{rcp3}
\eea
The 4D volume (the area of the boundary surface) is given by
\bea
A=\int\sqrt{\mbox{det}\gh_{ab}}~d^4p=\int\sqrt{1+\frac{\ptil^4}{(\frac{d\ptil}{dy})^2}}
~\frac{d\ptil}{\ptil^5}\nn
=\int_{1/\La}^l\sqrt{\left(\frac{\ptil'}{\ptil^2}\right)^2+1}~\ptil^{-3}dy\com\q \ptil'\equiv \frac{d\ptil}{dy}
\com
\label{rcp4}
\eea
where the $S^3$ property: $\ptil=1/\sqrt{x^2}=r(y)^{-1}=\ptil(y)$ is used. 
$\La$ is the UV cutoff for the 4D momentum integral. 
The minimal area principle gives the equation, 
\bea
\ptil(y)\ra \ptil(y)+\del\ptil(y)\com\q \del\Acal=0\com\nn
3+\frac{(\ptil''\ptil-2\ptil'^2)/\ptil^4}{1+(\ptil'/\ptil^2)^2}=0\q \mbox{or}\q
\frac{d^2\ptil}{dy^2}+\frac{1}{\ptil}\left(\frac{d\ptil}{dy}\right)^2+3\ptil^3=0
\pr
\label{rcp5}
\eea
This equation is the same as Eq.(\ref{surf2}) expressed by $r(=1/\ptil)$.

\subsection{Numerical confirmation of the relation between weight function 
and minimal surface curve}
The dominant configuration (path, $\ptil=\ptil(y)$) for Casimir energy Eq.(\ref{uncert1}) is 
given by an ordinary variation method (minimal ``action" principle).

\bea
y\ra y+\del y\com\nn
\del \{ W(\ptil(y),y)\ptil^3(y)F(\ptil(y),y)\}
=\del y \left(\frac{d\ptil}{dy}\frac{\pl}{\pl\ptil}+\frac{\pl}{\pl y}\right)
(\ptil^3W(\ptil,y)F(\ptil,y))=0,\nn
\frac{d\ptil}{dy}=\frac{-\frac{\pl\ln(WF)}{\pl y}}{\frac{3}{\ptil}+\frac{\pl\ln (WF)}{\pl\ptil}}
\com\q
\label{WeiGeo1}
\eea
where $W(\ptil,y)$ is the weight function `practically' introduced in \S~\ref{uncert} and 
is properly defined in \S~\ref{uncertb} using 5D coordinates $(r,y)$. 
In this subsection, we define $W(\ptil,y)$ using $(\ptil,y)$. 
$F(\ptil,y)$ is given in Eq.(\ref{UIreg2}). 
The coordinate version of the above result is given in the text (Eq.(\ref{uncert3})). 
We can also obtain $d^2\ptil/dy^2$.
\bea
\Wcal_y\equiv \pl_y\ln(WF)\com\q \Wcal_\ptil\equiv \pl_\ptil\ln(WF)\com\nn
\frac{d\ptil}{dy}=-\frac{\Wcal_y}{3\ptil^{-1}+\Wcal_\ptil}
\equiv \Vcal_1(W,\pl_\ptil W,\pl_y W; \ptil,y)\com\nn
\frac{d^2\ptil}{dy^2}=-\frac{\pl_y\Wcal_y}{3\ptil^{-1}+\Wcal_\ptil}
+\frac{\Wcal_y (\pl_\ptil\Wcal_y+\pl_y\Wcal_\ptil)}{(3\ptil^{-1}+\Wcal_\ptil)^2} 
-\frac{\Wcal_y^2 (-3\ptil^{-2}+\pl_\ptil\Wcal_\ptil)}{(3\ptil^{-1}+\Wcal_\ptil)^3}\nn
\equiv\Vcal_2(W,\pl_\ptil W,\pl_y W,\pl^2_\ptil W,\pl_y\pl_\ptil W,\pl^2_y W; \ptil,y)
\pr
\label{WeiGeo2}
\eea
Note that the RHSs of the expressions $d\ptil/dy$ and $d^2\ptil/dy^2$ above are functions of
$\ptil$ and $y$. If we consider that $W(\ptil,y)$ is unknown, by inserting the above ones
in Eq.(\ref{rcp5}), we obtain a partial differential equation for $W(\ptil,y)$ involving 
up to 2nd derivative. 
We call it the ``W-defining equation". 
\bea
\mbox{W-defining Equation}:\q
\Vcal_2+\frac{1}{\ptil}\Vcal_1^2+3\ptil^3=0
\com
\label{WeiGeo3}
\eea
where $\Vcal_1$ and $\Vcal_2$ are defined in Eq.(\ref{WeiGeo2}). 
We consider that the 2nd-derivative differential equation defines the weight function $W(\ptil,y)$. 
It is difficult to solve it. 
Here, we are content with a numerical consistency check.

If we take some example of $W(\ptil,y)$ appearing in Eq.(\ref{uncert1}), 
the dominant configuration $\ptil_W=\ptil_W(y)$ is graphically shown by the valley-bottom line 
of Casimir energy integrand $W(\ptil,y)\ptil^3F(\ptil,y)$, (see Figs.~\ref{GraphW1L10m1}--\ref{GraphW6L10m05}). 
On the other hand, there exists another path $\ptil_g(y)$ that is the minimal surface curve 
of the bulk geometry, that is, Eq.(\ref{surf2}) or (\ref{rcp5}). $\ptil_g(y)$ is  determined by the 5D metric, and is completely 
independent of both the weight $W$ and the model $F$.  
\footnote{
Precisely, 
the boundary conditions at $y=0$ and $y=l$ 
(boundary values:\ $\ptil(0)$ and $\ptil(l)$) 
are also necessary. 
} 
The $W$-defining equation above represents the equality 
$\ptil_W(y)=\ptil_g(y)$. 
We can numerically compare $\ptil_W(y)$ and $\ptil_g(y)$. 
%
\begin{figure}
\begin{center}
\includegraphics[height=8cm]{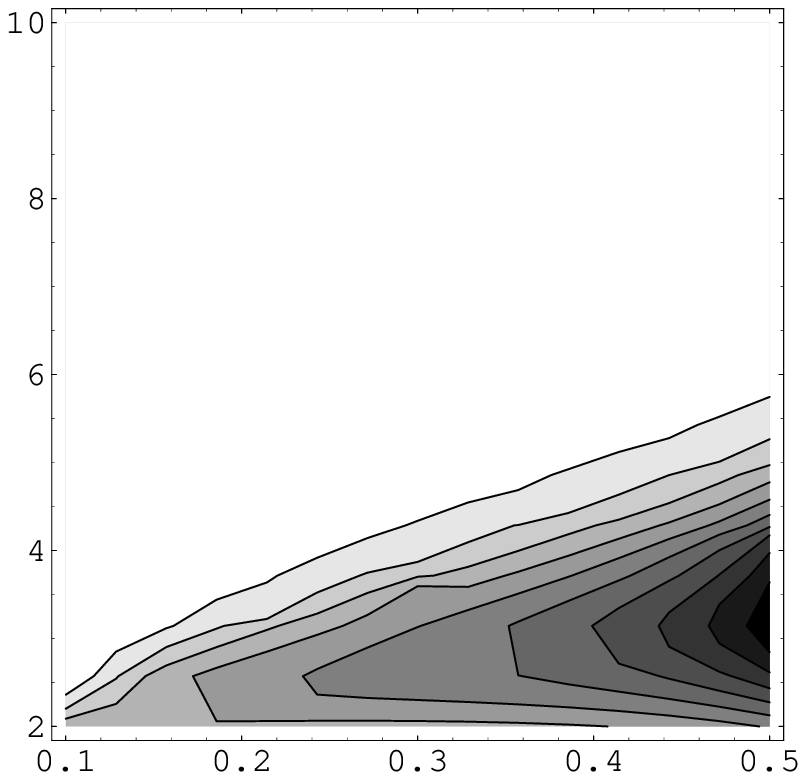}
\end{center}
\caption{
Contour of $\ptil^3W_6(\ptil,y)F(\ptil,y)$ (parabolic suppression 2, Fig.~\ref{GraphW6L10m05}). 
$\La=10,\ l=0.5$. 
Horizontal axis: $1.001/\La\leq y\leq 0.99999 l$;\ vertical axis: $1/l\leq \ptil\leq \La$. 
}
\label{ContW6La10m05}
\end{figure}
%
\begin{figure}
\begin{center}
\includegraphics[height=8cm]{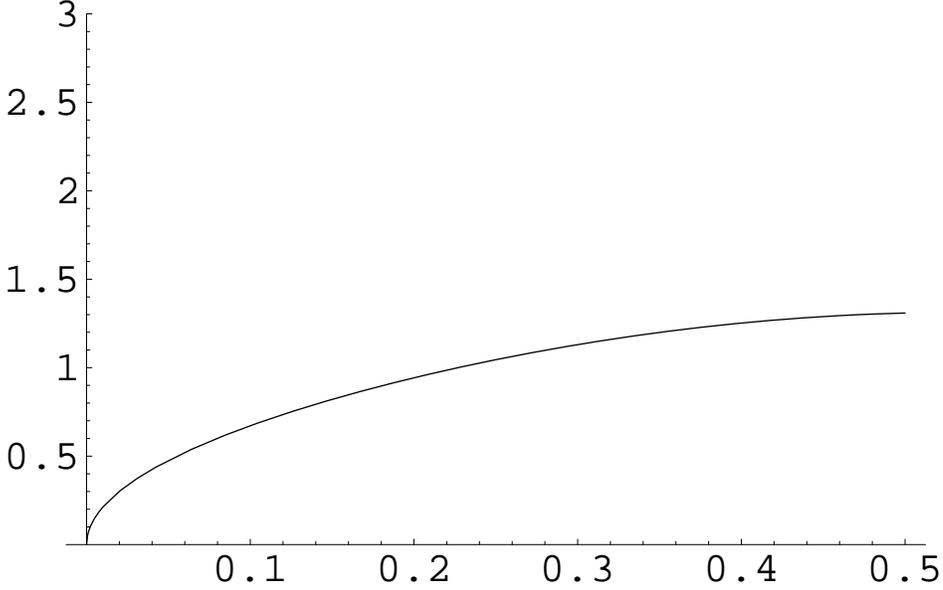}
\end{center}
\caption{
Minimal surface curve  $1/r_-(y), C=5.1215, C'=1.068$ in Eq.(\ref{A6}).  
Horizontal axis: $0\leq y\leq 0.5$;\ vertical axis: $0\leq 1/r_-\leq 3$. 
}
\label{GeodC5p1CP1p1}
\end{figure}
We show, in Fig.~\ref{ContW6La10m05}, the contour of Fig.~\ref{GraphW6L10m05}
(parabolic suppression W6). We can see the valley-bottom line as 
$\ptil_W(y)\approx 4.3\sqrt{y}$. In Fig.~\ref{GeodC5p1CP1p1}, we show 
the minimal surface curve $\ptil_g(y)=1/r_g(y)$, which is the $r_-(y)$ solution with 
$C=5.1215$ and $C'=1.068$ in  Eq.(\ref{A6}) of Appendix A.2. The two graphs are similar, at least
in shape and magnitude order. For other weights, we confirm a 
similar situation.
\section{Numerical evaluation of scaling laws: $E_{Cas}$ (\ref{UIreg5}), $E_{Cas}^{RS}$ (\ref{surfM1}), 
and $E_{Cas}^W$ (\ref{uncert1b})\label{appC}}
In the text, (regularized) Casimir energy is numerically calculated in three ways:\ 
1) original version (rectangle-region integral), 2) restricted-region integral ( Randall-Sundrum type ), 
3) weighted version. The final expressions show the {\it scaling} behaviors about the boundary 
(extra-space) parameter $l$ and the 4D momentum cutoff $\La$. The results are crucial 
for the present conclusion. Hence, we expalin here how the numerical results are obtained.

\begin{figure}
\begin{center}
\includegraphics[height=8cm]{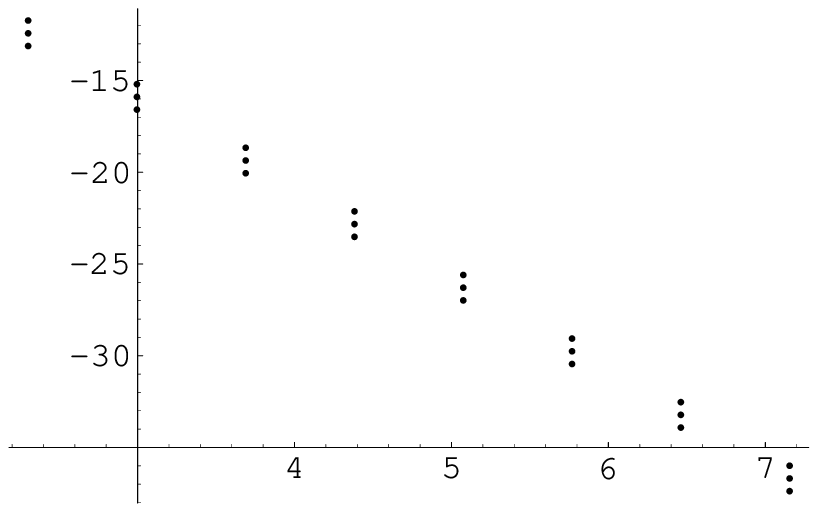}
\end{center}
\caption{
Casimir Energy  $E_{Cas}$ of (\ref{AppC1}) for various $(\La,l)$ values.  
Horizontal axis: $\ln \La$ ($\La=10,20,40,\cdots,1280$);\ vertical axis: $-\ln (|2^3\pi^2E_{Cas}|)$.  
The results are grouped into three lines. The values placed on the top, middle, and bottom lines 
correspond to $l= 10, 20,$ and $40$ respectively. 
}
\label{CasEneOrg}
\end{figure}

First, let us take the unweighted case with the rectangle integral region (original form) of 
Casimir energy (\ref{UIreg2}). 
\bea
2^3\pi^2E_{Cas}(\La,l)=\int_{1/l}^{\La}d\ptil\int_{1/\La}^ldy~\ptil^3 F(\ptil,y)
\com
\label{AppC1}
\eea
where $\ptil^3 F(\ptil,y)$ is explicitly given in Eq.(\ref{UIreg4}). The integral region 
is graphically shown in Fig.~\ref{ypINTregion} as a rectangle ($\ep=1/\La, \m=1/l$). 
The graphs of the integrand of (\ref{AppC1}), $\ptil^3F(\ptil,y)$, are shown for 
$(l,\La)=(1,10) \mbox{ [Fig.~\ref{p3F10La}]}, (1,100) \mbox{ [Fig.~\ref{p3F100La}]}, 
(1,1000) \mbox{ [Fig.~\ref{p3F1000La}]}, $
in the text. From the behaviors we can expect that $E_{Cas}(\La,l)$, (\ref{AppC1}), leadingly behaves as $l\La^5$, because 
the depth of the valleys, shown in Figs.~\ref{p3F10La}--\ref{p3F1000La}, is proportional to $\La^4$ and the graph-behaviors 
are monotonic along the y-axis (except near the boundaries $y=1/\La$ and $l$). 
It is confirmed by directly evaluating Eq.(\ref{AppC1}) 
numerically (the numerical integral in Ref.\citen{Mathema}). We plot the numerical results for various $\La$ and $l$ values in Fig.~\ref{CasEneOrg}. 
From the straight-line behavior we can safely fit the curve as 
$2^3\pi^2E_{Cas}=l\La^5(a_5+b_5\ln(l\La))$. The best fit is given by 
(manipulating numerical data in Ref.\citen{Mathema})
\bea
2^3\pi^2E_{Cas}(\La,l)=-0.1249~l\La^5 - (1.41, 0.706, 0.353)\times 10^{-5}~l\La^5\ln(l\La)
\pr
\label{AppC2}
\eea
The triplet results correspond to $l=10,20$ and $40$. 
The first term is firmly fixed (the number of significant figures (NSF) is 4), whereas the second term 
is unstable (the coefficients are proportional to $1/l$). 
The second one is numerically small compared with the first, and its determination
requires more careful treatment of small numbers. 
To determine it firmly, calculation using 
larger values of $\La$ and $l$ is necessary. 

\begin{figure}
\begin{center}
\includegraphics[height=8cm]{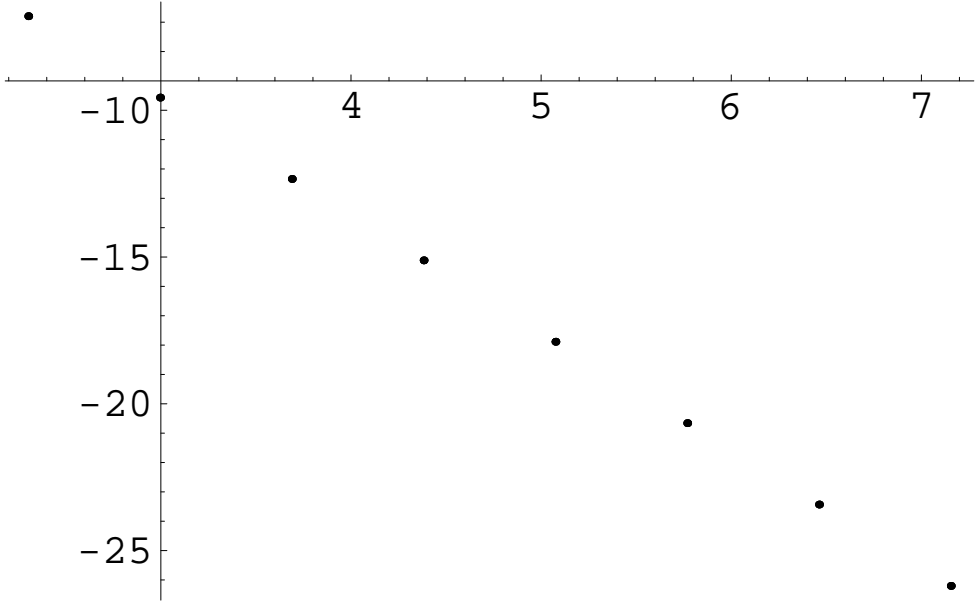}
\end{center}
\caption{
Casimir Energy  $E_{Cas}^{RS}$ of (\ref{AppC3}) for various values $(\La,l)$.  
Horizontal axis: $\ln \La$ ($\La=10,20,40,\cdots,1280$);\ vertical axis: $-\ln (|2^3\pi^2E_{Cas}^{RS}|)$.  
The results are placed on a straight line for different $\La$ and overlap (within the presented dots) for  
three different values of $l= 10, 20,$ and $40$. 
}
\label{CasEneRS}
\end{figure}

In the restricted region case, Eq.(\ref{surfM1}), we do the numerical integral of 
the following expression, 
\bea
2^3\pi^2E^{RS}_{Cas}=
\int_{1/l}^{\La}dq\int_{1/\La}^{1/q}dy~q^3 F(q,y)
=\int_{1/\La}^{l}du\int_{1/l}^{1/u}d\ptil~\ptil^3 F(\ptil,u)
\pr
\label{AppC3}
\eea
We plot the results in Fig.~\ref{CasEneRS} for various values of $\La$ and $l$. 
From the straight line behavior, we can safely fit the curve as 
$2^3\pi^2E_{Cas}^{RS}=\La^4(a_4+b_4\ln(l\La))$. The best fit is given by 
\bea
2^3\pi^2E_{Cas}^{RS}(\La,l)=\nn
-8.93814\times 10^{-2}~\La^4 + (+7.73, -4.83, +5.00)\times 10^{-10}~\La^4\ln(l\La)
\pr
\label{AppC4}
\eea
The triplet results corresponds to $l=10,20,$ and $40$. 
The first term is firmly fixed (the NSF is 6), whereas the second term 
is unstable. 
The second one is numerically very small compared with the first, and 
we may say that 
the second term vanishes 
within the present numerical precision. 

\begin{figure}
\begin{center}
\includegraphics[height=8cm]{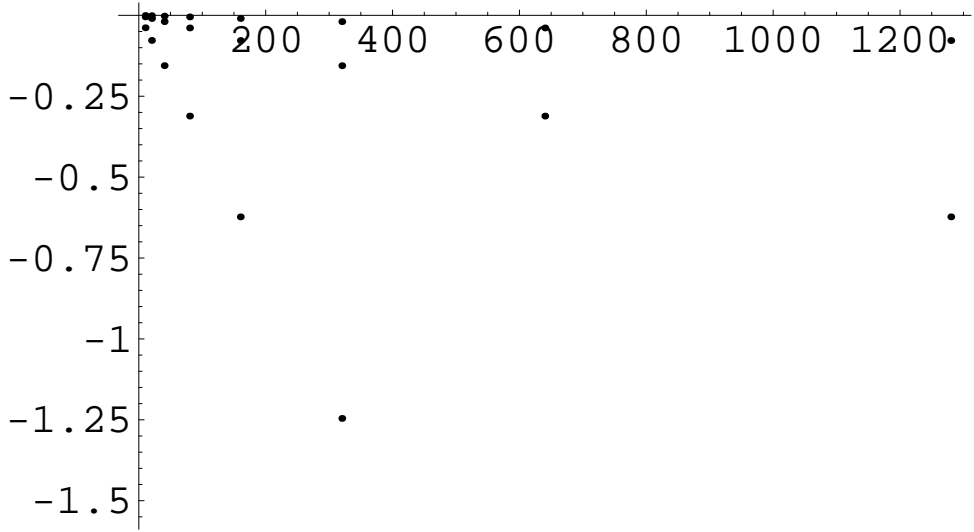}
\end{center}
\caption{
Casimir Energy  $E_{Cas}^{W_1}$ of (\ref{AppC5}) for various $(\La,l)$ values.  
Horizontal axis: $\La$ ($\La=10,20,40,\cdots,1280$);\ vertical axis: $2^3\pi^2E_{Cas}^{W_1}\times N_1$.  
The results are grouped into three lines. The values placed on the top, middle, and bottom lines 
correspond to $l= 40, 20,$ and $10$ respectively. 
}
\label{CasEneW1}
\end{figure}

Finally, we explain the weighted case in Eq.(\ref{uncert1}) taking the elliptic type $W_1$ as an example. 
\bea
2^3\pi^2E^{W_1}_{Cas}(\La,l)=\int_{1/l}^\La d\ptil\int_{1/\La}^ldy~\ptil^3 W_1(\ptil,y)F(\ptil,y)\com\nn
W_1(\ptil,y)=\frac{1}{N_1}e^{-\frac{l^2}{2}\ptil^2-\frac{1}{2l^2}y^2}\com\q
N_1=\frac{1.557}{8\pi^2}
\com
\label{AppC5}
\eea
where the UV cutoff $\La$ and IR cutoff $l$ are introduced to see the scaling behavior. In Fig.~\ref{CasEneW1}, we show 
the numerical results of $E^{W_1}_{Cas}(\La,l)\times N_1$ for different $\La$ and $l$ values. 
(Note that the axes of 
Fig.~\ref{CasEneW1} are on a linear scale, not on a log scale.)
From the straight line behavior of Fig.~\ref{CasEneW1}, we can safely fit the curve as 
$2^3\pi^2E_{Cas}^{W1}\times N_1=(\La/l)(a_1+b_1\ln(l\La))$. The best fit is given by 
\bea
2^3\pi^2E_{Cas}^{W_1}(\La,l)\times N_1=-(3.892,3.894,3.894)~\frac{\La}{l^3} + (-0.221, 1.70, 1.76)\times 10^{-4}~\frac{\La\ln(l\La)}{l^3}.
\label{AppC6}
\eea
The triplet results correspond to $l=10,20,$ and $40$. 
The first term is firmly fixed (the NSF is 3 at least), whereas the second term 
is unstable. 
This data shows that the coefficients become stable as $l$ increase. 
As for other types of $W$, the best fit scaling behaviors are listed in Eq.(\ref{uncert1b}) of the text. 
The behaviors taking $W_2$ and $W_3$ are different from the others. This is because the normalization factors 
$N_2$ and $N_3$ are divergent. The leading values, except $W_2$ and $W_3$, do not so much depend on 
the choice of $W$.  


\begin{thebibliography}{99}
\bibitem{Kal21} 
Th. Kaluza, Sitzungsberichte der K. Preussischen (Akademite der 
Wissenschaften, Berlin, 1921), p.966. 
\bibitem{Klein26} 
O. Klein, Z. Phys. \textbf{37} (1926),895.
\bibitem{AC83} 
T. Appelquist and A. Chodos, \PRD{28,1983,772}. 
\\
T. Appelquist and A. Chodos, \PRL{50,1983,141}. 
\bibitem{CasBMM01}  
M. Bordag, U. Mohideen and V. M. Mostepanenko, \PRP{353,2001,1} 
, quant-ph/0106045. 
\bibitem{ACGprl01}  
N. Arkani-Hamed, A. G. Cohen and H. Georgi, \PRL{86,2001,4757} 
, hep-th/0104005.
\bibitem{HPWpr01}  
C. T. Hill, S. Pokorski and J. Wang, \PRD{64,2001,105005} 
, hep-th/0104035.
\bibitem{BLS03}  
F. Bauer, M. Lindner and G. Seidl, \JHEP{05,2004,026} 
, hep-th/0309200.
\bibitem{RST05}  
L. Randall, M. D. Schwartz and S. Thambyahpillai, \JHEP{10,2005,110} 
, hep-th/0507102.
\bibitem{RS01}   
L. Randall and M. D. Schwartz, \JHEP{11,2001,003} 
, hep-th/0108114.
\bibitem{IM0703}  
S. Ichinose and A. Murayama, \PRD{76,2007,065008} 
, hep-th/0703228. 
\bibitem{FMMRnp9804}  
D. Z. Freedman, S. D. Mathur, A. Matusis and L. Rastelli, \NPB{546,1999,96} 
, hep-th/9804058.
\bibitem{HSjh9806}   
M. Henningson and K. Skenderis, \JHEP{07,1998,023} 
, hep-th/9806087.
\bibitem{DZatmp9810}  
J. Distler and F. Zamora, Adv.~Theor.~Math.~Phys. \textbf{2} (1999), 1405, hep-th/9810206.
\bibitem{GPPZjh9810}  
L. Girardello, M. Petrini, M. Porrati and A. Zaffaroni, \JHEP{12,1998,022} 
, hep-th/9810126.
\bibitem{PSpl9903}  
M. Porrati and A. Starinets, \PLB{454,1999,77} 
, hep-th/990385.
\bibitem{FGPWatmp9904}  
D. Z. Freedman, S. S. Gubser, K. Pilch and N. P. Warner, Adv.~Theor.~Math.~Phys.~\textbf{3} (1999) 363, hep-th/9904017.
\bibitem{Nambu70}  
Y. Nambu, ``Duality and Hydrodynamics", Lecture Notes at the Copenhagen Symposium (unpublished), 1970. 
\bibitem{Goto71}  
T. Goto, \PTP{46,1971,1560}. 
\bibitem{Pol81}  
A.M. Polyakov, \PLB{103,1981,207}. 
\bibitem{SI85PLB} 
S. Ichinose, \PLB{152,1985,56}. 
\bibitem{Schwinger51}  
J. Schwinger, Phys.~Rev.~\textbf{82} (1951) 664.
\bibitem{Dirac39}  
P. A. M. Dirac, {\it The Principles of Quantum Mechanics}, 4th Edition (Oxford Univ. Press, Oxford, 1958). 
\bibitem{MP9712}  
E. A. Mirabelli and M. E. Peskin, \PRD{58,1998,065002} 
, hep-th/9712214.
\bibitem{IM05NPB} 
S. Ichinose and A. Murayama, \NPB{710,2005,255} 
, hep-th/0401011.
\bibitem{SI07Nara}  
S. Ichinose, ``Casimir and Vacuum Energy of 5D Warped System and Sphere Lattice Regularization", 
Proc. of VIII Asia-Pacific Int. Conf. on Gravitation and Astrophysics 
,ed. M. Kenmoku and M. Sasaki 
(ICGA8, Aug.~29-Sep.~1,2007, Nara Women's Univ., Japan), 
p36, arXiv:/0712.4043.
\bibitem{SI07Osaka}  
S. Ichinose, \IJMP{23A,2008,2245} 
; 
Proc. of the Int. Conf. on Prog. of String Theory and Quantum Field Theory 
,ed. K. Fujiwara et al (Dec.~7-10,~2007,~Osaka City Univ., Japan), arXiv:/0804.0945.
\bibitem{SI08Singap}  
S. Ichinose, ``Casimir Energy of AdS5 Electromagnetism and Cosmological Constant Problem", 
Talk at Int. Conf. on ``Particle Physics, Astrophysics and Quantum Field Theory: 75 Years 
since Solvay" 
(PAQFT08,Nov.~27-Nov.~29, 2008, Nanyang Executive Center, Singapore), to appear in the 
conference proceedings, arXiv:/0903.4971.
\bibitem{SI0812}  
S. Ichinose, ``Casimir Energy of 5D Warped System and Sphere Lattice Regularization", 
arXiv:/0812.1263(hep-th).
\bibitem{Fey72}  
R. P. Feynman, {\it Statistical Mechanics},  (W.~A.~Benjamin, Inc., Massachusetts, 1972). 
\bibitem{Yoneya87}  
T. Yoneya, ``Duality and Indeterminancy Principle in String Theory" in {\it Wandering in the Fields}, 
ed. K. Kawarabayashi and A. Ukawa (World Scientific, 1987), p.~419.\\
T. Yoneya, ``String Theory and Quantum Gravity" in {\it Quantum String Theory}, 
ed. N. Kawamoto and T. Kugo (Springer, 1988), p.~23.\\
T. Yoneya, \PTP{103,2000,1081}. 
\bibitem{Mathema} 
S. Wolfram, {\it The Mathematica Book}, 4th ed., (Wolfram Media/Cambridge University Press, 1999).
\end{thebibliography}
\end{document}